\documentclass[a4paper]{article}
\pdfoutput=1 % if your are submitting a pdflatex (i.e. if you have
\usepackage{jcappub} % for details on the use of the package, please
\usepackage{natbib,xcolor,amsmath,amssymb,bm,multirow}
\usepackage[T1]{fontenc} % if needed
\usepackage[utf8]{inputenc}
\newcommand{\neal}[1]{{\textcolor{red}{\bf [#1]}}}

\newcommand{\hSI}{h_{\rm SI}}
\newcommand{\sigmaT}{\sigma_{\rm T}}
\newcommand{\hMsun}{h^{-1}\,M_\odot}
\newcommand{\hGpc}{h^{-1}\, \mathrm{Gpc}}
\newcommand{\hMpc}{h^{-1}\, \mathrm{Mpc}}

\title{\boldmath Signatures of Self-Interacting dark matter on cluster density profile and subhalo distributions}

\author[a,b,c,1]{Arka Banerjee,\note{Corresponding author.}}
\author[a,b,c,2]{Susmita Adhikari,\note{Corresponding author.}}
\author[d]{Neal Dalal,}
\author[e,f]{Surhud More,}
\author[g,h,i]{Andrey Kravtsov}

\affiliation[a]{Kavli Institute for Particle Astrophysics and Cosmology, Stanford University, 452 Lomita Mall, Stanford, CA 94305, USA}
\affiliation[b]{Department of Physics, Stanford University, 382 Via Pueblo Mall, Stanford, CA 94305, USA}
\affiliation[c]{SLAC National Accelerator Laboratory, 2575 Sand Hill Road, Menlo Park, CA  94025, USA}
\affiliation[d]{Perimeter Institute for Theoretical Physics, 31 Caroline Street North, Waterloo, ON N2L 2Y5, Canada}
\affiliation[e]{Kavli Institute for the Physics and Mathematics of the Universe (Kavli IPMU, WPI), University
of Tokyo, Chiba 277-8582, Japan}
\affiliation[f]{The Inter-University Center for Astronomy and Astrophysics, Post bag 4, Ganeshkhind,
Pune, 411007, India}
\affiliation[g]{Department of Astronomy and Astrophysics, The University of
Chicago, Chicago, IL 60637, USA}
\affiliation[h]{ Kavli Institute for Cosmological Physics, The University of Chicago,
Chicago, IL 60637, USA}
\affiliation[i]{ Enrico Fermi Institute, The University of Chicago, Chicago, IL 60637,
USA}
\emailAdd{arkab@stanford.edu}
\emailAdd{susmita@stanford.edu}
\emailAdd{ndalal@perimeterinstitute.ca}
\emailAdd{surhud.more@ipmu.jp}
\emailAdd{kravtsov@uchicago.edu}
\abstract{Non-gravitational interactions between dark matter particles with strong scattering, but relatively small annihilation and dissipation, has been proposed to match various observables on cluster and group scales. In this paper, we present the results from large cosmological simulations which include the effects of different self-interaction scenarios. In particular we explore a model with the differential cross section that can depend on both the relative velocity of the interacting particles and the angle of scattering. We focus on how quantities, such as the stacked density profiles, subhalo counts and the splashback radius change as a function of different forms of self-interaction. We find that self-interactions not only affect the central region of the cluster, the effect well known from previous studies, but also significantly alter the distribution of subhalos and the density of particles out to the splashback radius. Our results suggest that current weak lensing data can already put constraints on the self-interaction cross-section that are only slightly weaker than the Bullet Cluster constraints ($\sigma/m \lesssim 2$ cm$^2/$g), and future lensing surveys should be able to tighten them even further making halo profiles on cluster scales a competitive probe for DM physics. 

}

\begin{document}
\maketitle
\flushbottom

%======================
\section{Introduction}
\label{sec:intro}
%======================

While the Cold Dark Matter (CDM) paradigm has been extremely successful in modeling observables such as the Cosmic Microwave Background radiation \cite{Aghanim:2018eyx,Hinshaw:2012aka}, and the clustering of galaxies and matter (e.g. \cite{Alam:2016hwk,Abbott:2017wau,Hildebrandt:2016iqg}) on large scales, there are certain features of this paradigm which seem to be in conflict with observables on smaller scales - scales comparable to, or smaller than the sizes of typical dark matter halos \cite{Klypin:1999uc,deBlok:2009sp,Kravtsov:2010,BoylanKolchin:2011de,Weinberg:2013aya,Bullock:2017,Tulin:2017ara}. One of the underlying assumptions of the $\Lambda$CDM paradigm is that dark matter particles do not interact with either the Standard Model particles, or with other dark matter particles, except by gravity. In recent years, there has been growing interest in exploring fundamental physics models for the nature of dark matter, and the non-gravitational interactions of dark matter expected from these models (e.g. \cite{Jungman:1995df,Bergstrom:2000pn,Bertone:2004pz,Hooper:2007qk,Dodelson:1993je,Abazajian:2001nj} and references therein).  While non-negligible interactions of dark matter particles with baryons are constrained by direct detection experiments, there are interesting regions in the parameter space of interactions of dark matter particles with other dark matter particles that are not yet ruled out, but can produce possibly observable signatures on exactly the scales where prediction of $\Lambda$CDM cosmologies are in tension with observables.  

Specifically, dark matter self-interactions with strong scattering, but relatively small annihilation and dissipation, have been proposed as a way to alleviate some of these apparent small-scale problems of $\Lambda$CDM cosmology \cite{Spergel:1999mh}. Such interactions can potentially explain
the flat or ``cored'' density distribution in the central regions of many dwarf galaxies \cite{deBlok:2009sp}, known as the core-cusp problem. 
The basic idea is that non-gravitational scattering allows particles to exchange energy with each other, transferring thermal energy from large radii to small radii and thereby heating up the dark matter in high density regions. This process increases the sizes of particle orbits in the inner regions, suppressing the central density and flattening the central density profile. These processes enable these self-interaction models explain the diversity seen in rotation curves of galaxies at fixed halo mass and stellar mass \cite{Kamada:2016euw,Ren:2018jpt}. It should be noted that baryonic effects can also potentially account for the distribution of galaxy rotation curves \cite{Onorbe:2015ija,Bullock:2017,Contigiani:2019}, but these models usually make different predictions for the correlation between stellar mass and central dark matter density \cite{Read:2018fxs}, than
those expected from self-interaction models (e.g. \cite{Sameie:2019zfo}). For these reasons, it remains worthwhile to explore the various effects of dark matter self-interactions through cosmological simulations.

In this paper, we will study DM scattering with (relatively) short-range interactions that can be modeled as contact interactions in simulations, see  \cite{Nusser:2004qu,Kesden:2006zb,Kesden:2006vz} for examples of simulations with long-range self-interactions.  For extremely heavy mediators, the DM interaction can be treated as hard-sphere scattering in which the differential cross section is isotropic and independent of relative velocity, $d\sigma/d\Omega={\rm const}$. Most cosmological simulations with SIDM have focused on this scenario.  For lighter mediators, however, the differential cross section can be more complicated, with significant anisotropy and velocity dependence.  This change in the form of the cross section can potentially produce different macroscopic effects compared to the standard hard-sphere scenario.  This has motivated studies investigating the effects of anisotropy and velocity dependence in the differential cross section on various observables \cite{Kahlhoefer:2013dca,Kummer:2017bhr,Robertson:2016qef}. One example of this is drag vs.\ evaporation of DM subhalos orbiting within larger hosts.  A subhalo's DM particles can exchange energy and momentum with the ambient DM particles within its host halo, and these interactions can act similarly to interactions of baryon particles that produce ram pressure stripping, and produce a net drag from the cumulative effect of scattering events that do not unbind subhalo particles (analogous to dynamical friction).  Both of these effects arise in SIDM simulations, but \cite{Kahlhoefer:2013dca} speculated that a significant anisotropy in the differential cross section could enhance the effects of the ``drag'' force compared to the effect of mass evaporation. One effect of drag is the displacement of the light profile of the galaxy from the center of mass of the subhalo \cite{Harvey:2015hha,Massey:2015dkw}, which can potentially be measured through a combination of optical data and lensing measurements. Other effects include characteristic changes in the shape of disk galaxies due to this non-zero displacement between the subhalo's center and the plane of the galaxy \cite{Secco:2017vyp}.

While previous work has focused on evolution of exquisitely resolved cores,  evaporation of subhalos, or mergers of massive clusters \cite{GnedinOstriker:2001,Vogelsberger:2012ku,Rocha:2012jg,Peter:2012jh,Vogelsberger:2015gpr,Kim:2016ujt,Brinckmann:2017uve,Robertson:2016xjh}, in this paper we focus on large cosmological boxes to look at statistical samples of cluster mass halos and their subhalos that cover the comprehensive range of histories and environments in the universe. Recently, a feature in the outskirts of the density profile of halos, the splashback radius \cite{Diemer:2014xya, Adhikari:2014lna, More:2015ufa, Shi:2016lwp}, has emerged as a probe for physics at the interface of galaxy formation \cite{Adhikari:2016gjw,Snaith:2017,Mansfield:2019ter} and cosmology \cite{Adhikari:2018izo, Adhikari:2016gjw}.
\cite{Mansfield:2017, Diemer:2017ecy}.
This radius, which is the location at which the slope of the density profile of a dark matter halo reaches a minumum,  corresponds to the tail of largest orbital apocenters of the accreted matter completing its first orbit in the halo potential.
Observationally it has been measured in several studies \cite{More:2016vgs, Baxter:2017csy, Chang:2017hjt, Shin:2018pic,Contigiani:2019} using the stacked  galaxy distribution and weak lensing around clusters in large galaxy surveys.  Models of self-interaction that produce a cumulative drag  can cause subhalos to lose energy and move splashback inwards \cite{More:2015ufa}.

In this paper, therefore, we investigate the effects of various forms of the self-interaction cross sections on the particle and subhalo density profiles of the outer regions of massive clusters and location of the splashback radius using a suite of cosmological simulations of relatively large volumes. The plan of the paper is as follows. In Section \ref{sec:method}, we discuss the method that we adopt to include the self-interactions into cosmological simulations. In Section \ref{sec:sim_details}, we discuss the specifications of the simulations and the fiducial cosmology we adopt. In Section \ref{sec:convergence_tests} we present the results of various convergence tests to show that our simulations deliver robust results in the regions of interest. In Section \ref{sec:results}, we present the actual results on the impact of different forms of self-interactions on various observables. Finally, in Section \ref{sec:summary}, we summarize our findings and discuss the prospect of constraining the self-interaction cross section of dark matter by combining all the observables considered here.

%=============================================
\section{Implementation of self-interactions}
\label{sec:method}
%=============================================

We model elastic self-interactions of dark matter particles in $N$-body simulations by implementing a method very similar to the one outlined in \cite{Rocha:2012jg}. Specifically, we implement these extra non-gravitational interactions in an optimized version of the  
\textsc{Gadget-2} $N$-body code. In a nutshell, whenever two simulation particles are within some distance (in simulations units) $\hSI$ in the simulation box, there is a finite probability for them to interact. As we will discuss in detail in this section, this probability may depend on the relative velocities of the interacting particles. If the two particles do interact, the scattering angle of the event is drawn from the distribution described by the differential cross section of the interaction. This scattering angle can then be used to straightforwardly modify the velocities of the interacting particles.

Following \cite{Rocha:2012jg}, we assume that for a pair of particles, labeled $i$ and $j$, separated by $\delta \mathbf x_{ij}$, the overlap fraction is given by 
\begin{eqnarray}
\label{eq:weight}
g_{ij}(r) = \int _0 ^{\hSI}d^3\mathbf x' W \left( |\mathbf x'|,\hSI \right)W\left(|\mathbf x' + \delta \mathbf  x_{ij}|, \hSI\right) \, ,
\end{eqnarray}
where $W(r,h)$ is the spline kernel \cite{Monaghan1985}:
\begin{equation}
    W(r,h) = \frac{8}{\pi h^3}\begin{cases}
    1-6\left(\frac{r}{h}\right)^2+6\left(\frac{r}{h}\right)^3,  \quad  0\leq \frac{r}{h} \leq \frac 1 2 \\
    2\left(1-\frac r h\right)^3, \qquad \qquad \quad \frac 1 2 \leq \frac r h \leq 1 \\
    0, \qquad \qquad \qquad \qquad \qquad \frac r h >1 \,.
    \end{cases}
\end{equation}
Then the probability of interaction between the particles is given by 
\begin{equation}
\label{eq:prob}
P_{ij} = \frac {\sigma(v_{\rm rel})} {m}\, m_i \, v_{\rm {rel}}\, g_{ij}\Delta t\,.
\end{equation}
Here $\sigma/ m$ is the total cross section per unit mass, and $m_i$ is the mass of the simulation particle, $v_{\rm rel}$ is the relative velocity of the two particles, and $\Delta t$ is the time step over which the probability is calculated. Note that for velocity-dependent differential cross sections, $\sigma/m$ will itself depend on $v_{\rm rel}$, while for velocity-independent interactions, it is a fixed value irrespective of the relative velocity of the interacting particles. We make the assumption that we always have the masses of the two simulation particles to be the same, but this condition can be suitably relaxed. For the timestep, we always choose the smaller timestep of the pair of particles, since different particles can have different time steps in the integration scheme of \textsc{Gadget-2}.  We also point out that all quantities that enter the calculation of the probability in Eq. \ref{eq:prob} should be physical quantities, and so appropriate conversion factors are needed in the above equation to convert from \textsc{Gadget-2}'s internal velocity and time units to physical velocities and physical time.

As in \cite{Rocha:2012jg}, we use the native tree-walk of \textsc{Gadget-2} to compute these extra forces. We modify the tree walk slightly so that for every particle we open all nodes that can have a neighbor within the interaction radius $\hSI$. Note that for the normal tree walk to compute gravitational interactions in \textsc{Gadget-2}, only information about the positions of particles is exchanged if two neighbors reside on different processors. Since the extra interactions also involve the velocities of the particles, we modify the communication step so as to communicate particle velocities as well as positions to other processors, as required. 

There are two places in the above implementation where the actual form of the cross section enters. The first, as referred to above, is in the form of $\sigma/m$ in Eq. \ref{eq:prob}. For the velocity-independent interactions, whether isotropic or anisotropic, $\sigma$ represents the total cross section 
\begin{equation}
\label{eq:total_cross_section}
\sigma = \int \frac{{\rm d} \sigma}{\rm d \Omega} \rm d \Omega \,,
\end{equation}
and does not depend explicitly on the relative velocities of the two particles. On the other hand, for velocity dependent interactions, $\rm d \sigma/ \rm d \Omega$ depends on the relative velocity, and so $\sigma$ represents the angle-integrated total cross section at that velocity. Once the form of the differential cross section has been specified, we can compute the corresponding $\sigma$ which enters Eq. \ref{eq:prob}. We then generate a random number from a uniform distribution between $0$ and $1$ and if the random number is smaller than the value of $P_{ij}$, we scatter the two particles. Note that for this treatment to be consistent, the time steps have to be constrained so that the typical $P_{ij}$ are much smaller than $1$. 

The second place where the differential cross section is relevant is, of course, in the probability distribution of the scattering angles of interacting particles. We detail our procedure for choosing the scattering angle below for each of the cases studied in this paper. In the following, we always refer to the scattering angle of the collisions in the center-of-mass frame.

\subsection{Velocity-independent isotropic cross section}
\label{sec:isotropic}
This is the simplest form of elastic self-interactions in terms of the structure of the differential cross section, i.e. $d\rm \sigma / d\rm \Omega$ does not depend on either the relative velocity or the solid angle $\Omega$. This form has been studied extensively in the literature \cite{Rocha:2012jg,Peter:2012jh,Vogelsberger:2012ku}. The choice of the scattering angle for each pair of interacting simulation particles is especially simple in this case: we assume that the cosine of the scattering angle, $\cos \theta_{\rm scatter}$, is distributed uniformly between $-1$ and $1$. However, for identical particles, we only need to consider scattering angles between $0$ and $\pi/2$, i.e. for $0\leq \cos\theta_{\rm scatter} \leq 1$. This is because for indistinguishable particles a scattering event with $\theta>\pi/2$ is kinematically indistinguishable from $\theta \prime = \theta - \pi/2 $. 

In our implementation, therefore, we generate a random number from a uniform distribution between $0$ and $1$ for every interacting pair and assign the scattering angle. Throughout this paper, we will refer to the strength of the isotropic interactions with the value of total cross section per unit mass, $\sigma/m$. This is in contrast to the anisotropic self-interactions discussed below.

\subsection{Velocity independent anisotropic cross section}
\label{sec:anisotropic}
As noted above, when DM scattering becomes anisotropic, the cross section also generally acquires significant velocity dependence.  To disentangle the different effects of velocity dependence and anisotropy, we first consider anisotropic scattering without velocity dependence, where 
$d\rm \sigma/d\rm \Omega$ is independent of the relative velocity of the interacting particles but does depend on the solid angle $\Omega$. We specifically adopt the form for the differential cross section studied in \cite{Kahlhoefer:2013dca}:
\begin{equation}
\label{eq:anisotropic}
\frac{d\rm \sigma}{d\rm \Omega} \propto \frac{1+\cos^2 \theta}{1-\cos ^2\theta} \,.
\end{equation}
Note that as $\cos\theta \to  1$, the differential cross section diverges. Therefore, interactions are much more likely to lead to small scattering angles and therefore, small momentum exchange, rather than large deviations in the center of mass frame. This is in contrast to the isotropic case, where interactions with both small and large momentum exchange are equally likely. 

Eq. \ref{eq:anisotropic} also implies that the total integrated cross section from Eq. \ref{eq:total_cross_section} also diverges, and care has to be taken when trying to generate the distribution function from which the scattering angle is drawn. It also means that to compare different strengths of self-interaction, we instead use the momentum transfer cross section, as defined in \cite{Kahlhoefer:2013dca}:
\begin{equation}
\label{eq:sigma_t}
\sigmaT =   \int \frac{\rm d \sigma}{\rm d\Omega}(1-|\cos\theta|)\rm d \Omega \,.
\end{equation}
This cross section is finite and well-behaved for the differential cross section defined in Eq. \ref{eq:anisotropic}. Therefore, the choice of normalizing the strength of self-interactions is different in this case compared to the isotropic interactions in \S \ref{sec:isotropic}. Note that one can also define $\sigmaT$ for an isotropic cross-section. In fact, it is straightforward to show that $\sigma/m = 2\,$cm$^2$/g for the isotropic cross-section implies $\sigmaT/m = 1\,$cm$^2$/g. 

To handle the divergence in $\sigma$ while determining scattering angles in simulations, we introduce a cutoff $\epsilon$ such that the integral in Eq. \ref{eq:total_cross_section} only runs from $0$ to $(1-\epsilon)$. We then check for the convergence of the various moments of the distribution function with respect to the value of $\epsilon$. We find that for the choice of $\epsilon = 0.01$, the first and second moments of the distribution which control the momentum and energy exchange of the particles, are converged to within $1\%$.

Once the choice of $\epsilon$ is fixed, we generate the cumulative distribution function for $x=\cos\theta$:
\begin{equation}
{\rm CDF}(x = \cos \theta) = \frac{\displaystyle \int^{x}_0 \frac{1+x^2}{1-x^2}{\rm d}x}{\displaystyle \int ^{1-\epsilon}_{0}\frac{1+x^2}{1-x^2}{\rm d} x} \,.
\label{eq:cdf_vel_ind}
\end{equation}
Next, we use the transform method to generate scattering angle. Namely, we generate a random number $R$ from a uniform distribution between $0$ and $1$, and then interpolate to get the value of $x$ such that ${\rm CDF} (x) = R$, which gives us the scattering angle for the interaction. Note that, for the rest of the paper, the strength of the anisotropic self-interactions will be referred to by the value of $\sigmaT/m$, the momentum cross-section per unit mass. As mentioned in Section \ref{sec:isotropic}, isotropic self-interactions will instead be labeled by their total cross section per unit mass $\sigma/m$.

\subsection{Velocity and angle dependent cross section}
\label{sec:velocity_dep}

\begin{figure}
\begin{center}
\includegraphics[width=0.50\linewidth]{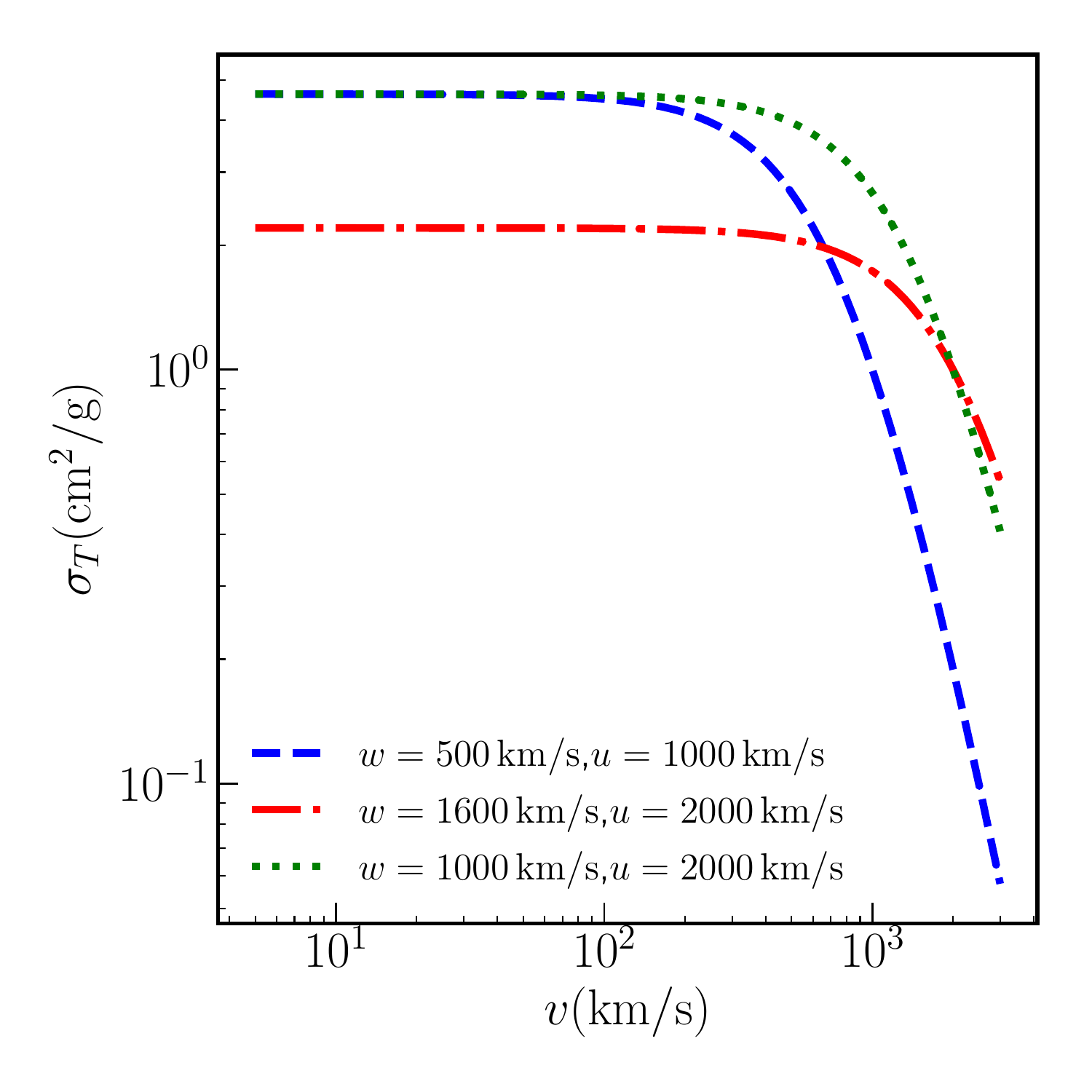}
\end{center}
\caption{Momentum transfer cross section for velocity dependent differential cross sections considered in this paper. At low velocities ($v\ll w$), the cross sections are isotropic and velocity-independent, while at high velocities $(v\gg w)$, the cross section scales as $v^{-4}$, along with an angular dependence given in Equation \ref{eq:vel_dep}.}
\label{fig:cross_section}
\end{figure}

The last self-interaction regime we consider is the most general elastic collisions, in which the differential cross section can depend on the relative velocity of the interacting particles, as well as the scattering angle. While the method we outline below can be generalized to any form of the cross section, for this paper we concentrate on the cross section given in \cite{Kummer:2017bhr,Ibe:2009mk} :
\begin{equation} 
\label{eq:vel_dep}
\frac{\rm d \sigma}{\rm d \Omega} = \frac{\displaystyle \sigma_0}
{\displaystyle 2\left[1+\frac{v^2}{w^2}\sin^2 \left(\frac{\theta}{2}\right)\right]^2} \,.
\end{equation}
Here $w$ acts as a velocity scale separating the isotropic cross-section regime ($v\ll w$) from the highly anisotropic Rutherford scattering regime ($v\gg w$). We will treat $w$ as a free parameter to see how this transition scale affects halo structure. The normalization factor $\sigma_0$ is defined following \cite{Kummer:2017bhr}: 
\begin{equation}
\label{eq:vel_dep_normalization}
\frac {\sigmaT(u)}{m} = 1\, {\rm cm^2/g} \,,
\end{equation}
where $u$ is some reference velocity. We also study how a change in the reference velocity at which we normalize the cross section affects the structure formation. We plot the momentum transfer cross section as a function of the interaction velocity for the different choices of $u$ and $w$ in Fig. \ref{fig:cross_section}. We choose $u$ and $w$ such that two of the interactions behave similarly in the low-velocity end of the cross section, with different behavior at the high velocity end. Another pair is chosen to have the same $u$, the normalization velocity, but a different transition scale between isotropic and anisotropic scattering.

To generate scattering angles, we follow a treatment similar to  \cite{Robertson:2016qef}  defining $500$ bins between $v_{\rm min} = 10 \, \rm{km/s}$ and $v_{\rm max} = 4000\, {\rm km /s}$, using the central value of each bin in the denominator of Eq. \ref{eq:vel_dep}, and then computing the required cumulative distribution for $x = \cos\theta$ in that bin exactly analogous to the method in \S \ref{sec:anisotropic}. Note that, throughout the rest of the paper, we will refer to different velocity-dependent interactions by using the values of $u$ and $w$ for that model.

%--------------------------------------------------%
\section{Simulation details}
\label{sec:sim_details}
%---------------------------------------------------%

All simulations presented in this paper adopt $\Lambda$CDM cosmology with $\Omega_{\rm m} = 0.3$, $\Omega_\Lambda = 0.7$, $A_s = 2.2\times 10^{-9}$, $n_s = 0.96$, and $H_0 = 70\,$km/s/Mpc, giving $\sigma_8=0.85$.  

We present two sets of simulations. The first set  simulates a volume of $(1\,\hGpc)^3$ with $1024^3$ dark matter particles. The gravitational spline softening length was chosen to be $0.015\, \hMpc$ which is $\sim 1/60$ of the mean inter-particle separation. 
The simulations were all run with the same initial conditions at $z=99$ generated using the publicly available \textsc{N-GenIC} code\cite{2015ascl.soft02003S}. We use the \textsc{Rockstar} halo finder \cite{Behroozi:2011ju} to identify halos in the simulations.
These simulations allow for a large enough sample of cluster-sized halos in the mass range $1-2\times 10^{14}\, \hMsun$ to obtain good statistics on quantities like the stacked profiles of substructures close to the outer boundaries of these cluster scale halos.

However, the simulation parameters mentioned above do not allow us to resolve dynamics of dark matter particles deep inside the average scale radius of these cluster mass halos. Therefore, to study the effects of dark matter self-interactions down to  smaller scale, in view of comparing to results from literature, we have also run simulations with volume $(500\,\hMpc)^3$ and gravitational softening of $0.0075\,\hMpc$, but using the same number of particles ($1024^3$) and the same
initial epoch ($z=99$).

\section{Code tests}
\label{sec:convergence_tests}
In the first part of this section, we show that our code recovers the features that have been reported in previous investigations of the SIDM models, specifically the formation of a core in the density profile of inner parts of dark matter halos. We then present the results of convergence tests to check if the observed stacked density profiles that we investigate in Sec. \ref{sec:results} are stable with respect to our choice of various parameters relevant for the SIDM interactions.

\subsection{Core formation and thermalization}
\label{sec:core_tests}

\begin{figure}[t]
    \centering
    \includegraphics[width=1.0\linewidth]{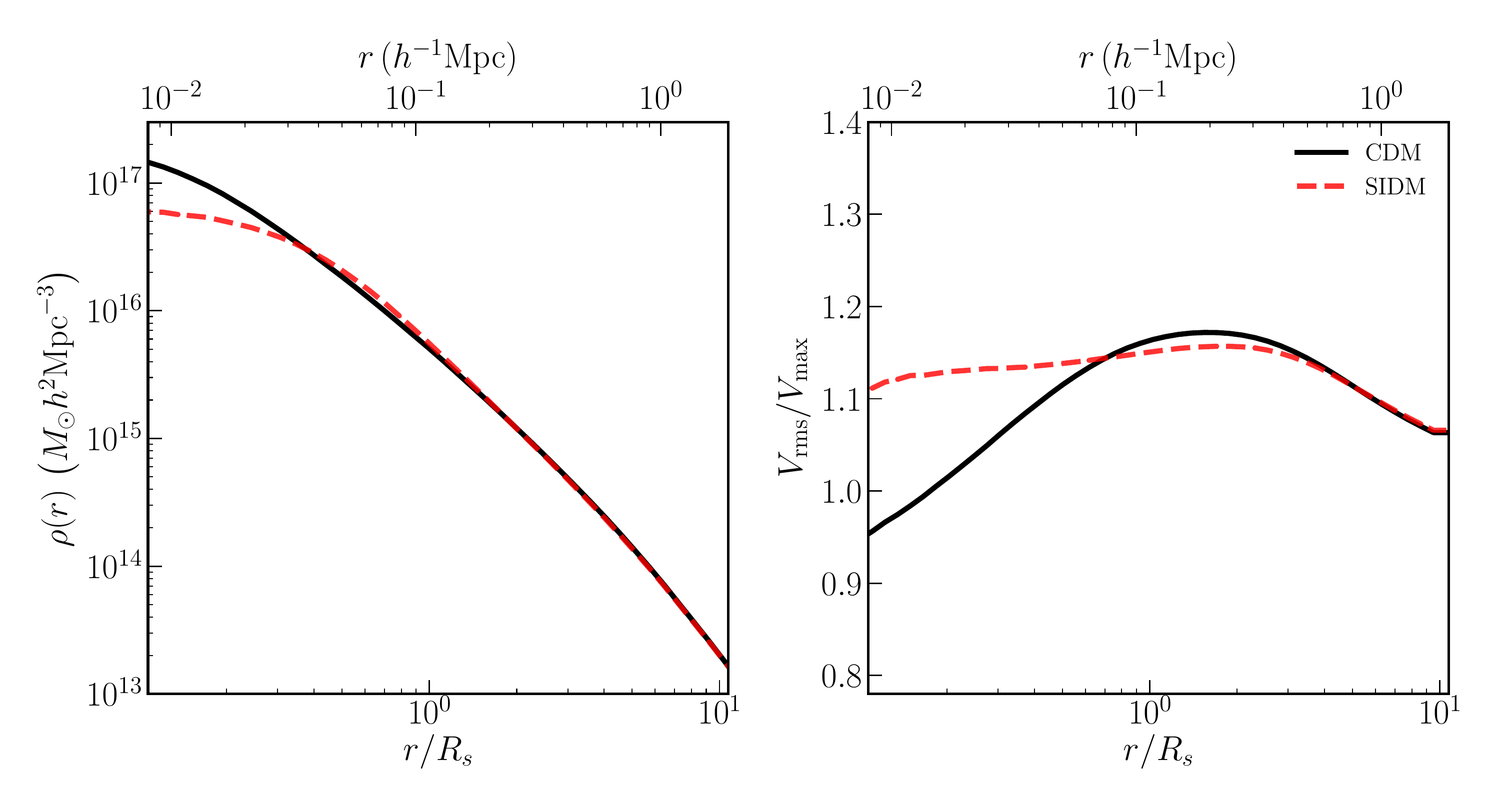}
    \caption{Stacked density profiles and stacked velocity dispersion of halos in the mass range $1\times10^{14}\,\hMsun$ to $2\times10^{14}\,\hMsun$ from the simulations with side $500\,\hMpc$ where the inner parts of these objects are relatively well-resolved. The black curves in each panel show the CDM density profile and dispersion, while the red curves are from simulations of SIDM with isotropic cross sections of $1\,$cm$^2/$g. The stacked density profile shows significant coring below the average scale radius of these halos ($0.14\, \hMpc$). The velocity dispersion profile shows thermalization  inside the scale radius halos in the presence of self-interactions.}
    \label{fig:thermalization}
\end{figure}

One of the most robust predictions of SIDM simulations with $\sigma/m \sim 1-10\,{\rm cm}^2/{\rm g}$ is the formation of a central core in the density profiles of halos \cite{Burkert:2000di,Balberg:2002ue,Ahn:2004xt,Koda:2011yb,Rocha:2012jg}. In the standard CDM scenario, the density profiles of halos is expected to be cuspy near the center, i.e. the density continues to increase as the radius decreases. However, the extra scattering between particles near the halo center in SIDM models can lead to observable deviations from the behavior seen in CDM simulations. This is especially true in regions where the densities and relative velocities are high enough for dark matter particles to have at least one interaction on average per dynamical time. 

This behavior has been extensively studied in many previous simulations of SIDM, e.g. \cite{Rocha:2012jg}, especially in the context of isotropic and velocity independent SIDM cross sections. While \cite{Rocha:2012jg} investigated the density profiles of individual halos, the large volumes of our simulations allow us to compute the stacked dark matter profiles around centers of halos in a fixed mass range. It has been observed that, for realistic SIDM cross-sections, the scale at which the core appears for halos of different masses is roughly determined by the scale radius. For the cluster sized halos in the mass range $1\times 10^{14} \hMsun$ to $2\times 10^{14} \hMsun$ that we consider, the average scale radius is $\sim 0.1 \hMpc$. To ensure that our simulations have sufficient resolution to get down to these scales, we consider the simulations run with box length $512 \hMpc$ and $1024^3$ particles, instead of the larger box with lower resolution. For a fair comparison to the previous results in the literature, we also consider the self-interactions to have a velocity independent and isotropic form with $\sigma/m=1\,$cm$^2/$g.

The results of this calculation are presented in the left panel of Fig. \ref{fig:thermalization}. The black curve shows the stacked density profile of halos in the given mass range from the CDM simulation, while the red curve shows the stacked density profile around halos of the same virial mass range from the SIDM simulation. For this value of the cross section, the outer profiles do not change appreciably, but below the scale radius, the density profile from the SIDM simulation shows a clear core compared to the steeper density profile from the CDM simulation, in agreement with previous work. 

As is well known in the literature, this central flattening arises from the thermalization of the inner regions of dark matter halos \cite{Rocha:2012jg,Peter:2012jh}. For CDM halos, the velocity dispersion increases from small radii out towards the scale radius $r_s$, so the inner parts of the halo are found to be ``colder'' than the outer regions. Self-interactions, on the other hand, transfer energy from larger radii to the inner regions, thereby heating the core. To check for this effect, we compute the velocity dispersion of particles in different radial bins around the centers of halos. Once again, the quantity we compute is stacked over all halos in the mass range present in the simulation volume. 

The results are presented in the right panel of Fig. \ref{fig:thermalization}. The black curve shows the dispersion profile from the CDM simulation, while the red curve shows the dispersion from the SIDM simulation. Instead of the absolute values of the radius from the halo center and the velocity dispersion, we rescale the distance by the scale radius of each CDM halo while the velocities are all rescaled by the maximum circular velocity of $V_{\rm max}$ of each CDM halo. For the calculation from the SIDM simulation, this requires the identification of the corresponding halo from the CDM simulation, and using its scale radius and $V_{\rm max}$ to re-scale. We find a clear signal of thermalization in the SIDM case, where the velocity dispersion remains roughly constant inside the scale radius. This is in marked contrast to the black curve, where the velocity dispersion decreases as one moves to smaller radius inside the halos. This is consistent with results from previous studies, e.g. \cite{Rocha:2012jg}.

\subsection{Convergence test for stacked density profiles}
\label{sec:conv_test}
 Next, we check how the stacked density profiles around cluster mass halos change as we change SIDM implementation parameters. Since the rest of our paper investigates the changes due to self-interactions on the outer parts of the halo profiles, we will also concentrate on these scales ($\gtrsim 0.1 \hMpc$). We fix the fiducial self-interaction scenario in the velocity independent case to be the anisotropic angular form of Equation \ref{eq:anisotropic} with $\sigmaT/m = 1\,$cm$^2$/g. For the velocity dependent cross sections, we choose $w=1600\,$km/s and $u=2000\,$km/s where $w$ and $u$ are defined as in Equations \ref{eq:vel_dep} and \ref{eq:vel_dep_normalization}. Since we do not need very high resolution for these tests, we use the simulation boxes with size $1 \hGpc$, which are also used for the rest of the analysis in this paper.
 
 \begin{figure}
\begin{center}
\includegraphics[width=0.45\linewidth]{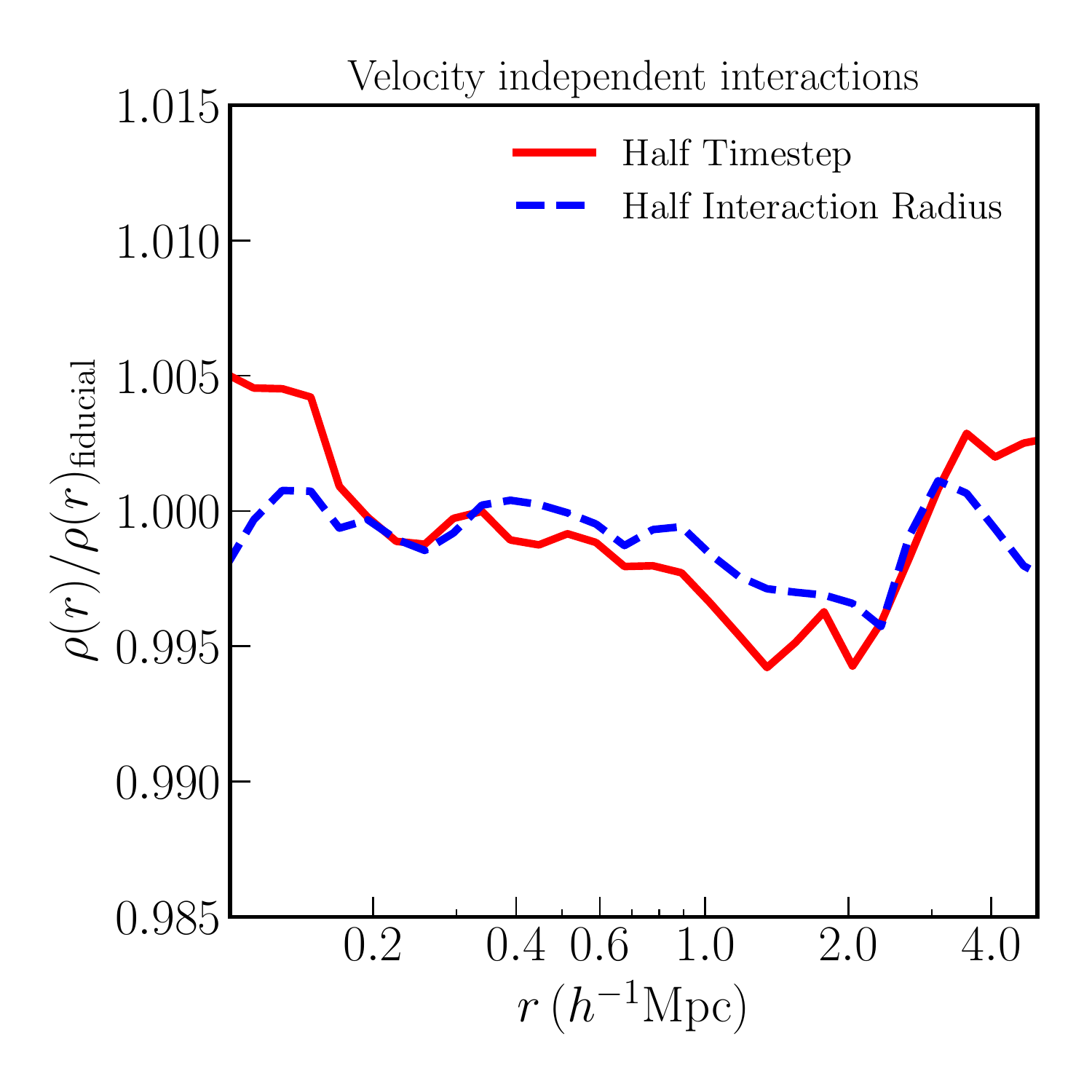}
\includegraphics[width=0.45\linewidth]{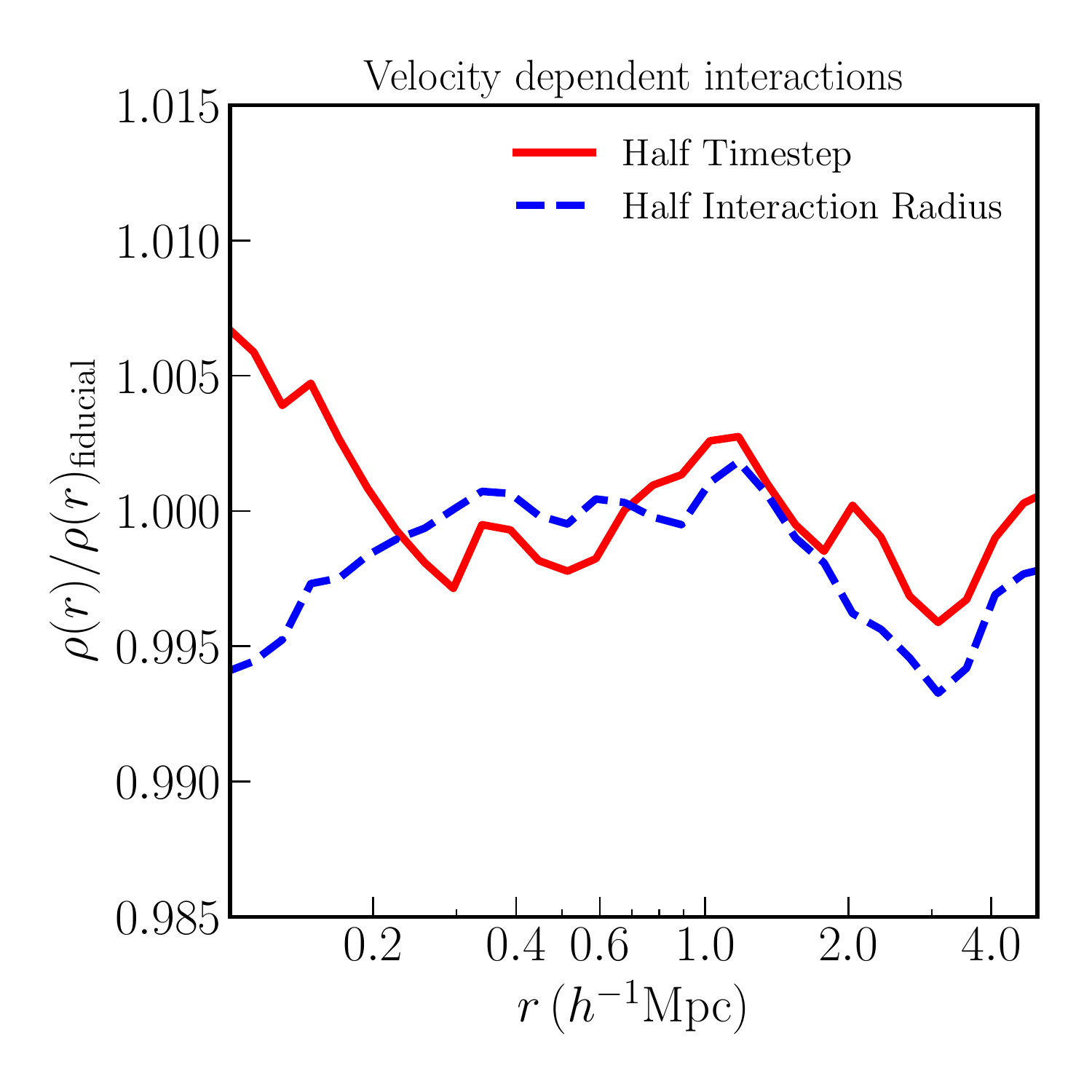}
\end{center}
\caption{Left: Ratio of stacked profiles of dark matter halos to the fiducial profile in the mass range $1\times 10^{14}\hMsun$ to $2\times 10^{14}\hMsun$ from the simulations with anisotropic $\sigmaT/m=1\,$cm$^2$/g. The red curve represents the ratio of the profiles when the timestep is halved from the fiducial value, while the blue curve represents the ratio to the fiducial when the interaction search radius is halved from the fiducial value. Right: Same as the left panel, but for velocity dependent interactions with $w=1600\,$km/s and $u=2000\,$km/s. The range of scales presented in this figure corresponds to the ones on which we will investigate the effects of self-interactions in the rest of the paper.}
\label{fig:conv_prof_ind}
\end{figure}

The first parameter that we investigate is the particle time step. The time step is relevant to the calculations of self-interactions because it appears explicitly in the interaction probabilities in Eq. \ref{eq:prob}. In the \textsc{Gadget-2} code, the time step of a particle is set by $\Delta t = \sqrt{2\eta\epsilon/|\mathbf a|}$, where $\mathbf a$ is the acceleration of the particle, $\epsilon$ is the gravitational softening length, and $\eta$ is a tunable parameter, with a fiducial value of $0.05$. In our convergence test, we change the value of $\eta$ to half that of the fiducial choice, and compute the change in the stacked density profile of cluster-sized objects.  We also test convergence  for the radius within which self-interactions are considered, $\hSI$. Our fiducial choice is $\hSI =  \epsilon$, where $\epsilon$ is the gravitational softening length, the choice made also by \cite{Rocha:2012jg}. For the convergence test, we change $\hSI$ to half of its fiducial choice, while keeping the gravitational softening length the same. This isolates the effect of the interaction radius on observables since the gravitational evolution should remain the same at a fixed gravitational softening length.

We present the results from these tests in Fig.  \ref{fig:conv_prof_ind}. We plot the results as ratios of the stacked density profiles around halos in the mass range $1-2\times10^{14} \hMsun$ to their value with the fiducial parameter choices. The left panel of the figure shows the results of the tests for the velocity independent cross section, while the right panel shows the results from the velocity dependent interactions. For both sets of interactions, we find that a change in either the search radius or the timesteps by a factor of two does not produce changes of larger than $1\%$ on any of the scales that we will focus on for the rest of the paper. In fact, for most of the radial bins, the actual changes are actually much smaller, with the largest differences appearing in the innermost bin where the profile is steeply rising.

We have also performed similar tests by comparing the relative change in the profile from the $(1 \hGpc)^3$ box to $(500 \hMpc)^3$ boxes, while using the same number of simulation particles, so that the mass of an individual particle goes down by a factor of $8$, and the spatial resolution goes up by a factor of $2$. Once again, we find that the change is consistent with the fiducial simulations at the sub-percent level.

We conclude that on the mass and length scales we are interested in, i.e., on the outskirts of cluster mass halos, the changes in various measured quantities change by at the most $\sim 1\%$ when our fiducial parameter choices are changed by a factor of $2$.
We can therefore expect that the results presented in the next section are robust to these choices. As will be shown, the different scenarios that we explore in the next section produce much larger changes than we see in these convergence tests, which means that we can be confident in interpreting the results in the next section as truly coming from the changes in the cross section of self-interactions, rather than our exact choice of parameters for implementing the self-interactions.

%-------------------------------------%
\section{Results}
\label{sec:results}
%-------------------------------------%
In this section, we present the main results of our investigations. A note on labeling conventions: whenever we refer to models of velocity-independent and isotropic self-interactions, we use the value of the total cross section per unit mass $\sigma/m$. For velocity-independent self-interactions with anisotropic differential cross section given by Equation \ref{eq:anisotropic}, we refer to each by the value of the momentum transfer cross section (defined by Equation \ref{eq:sigma_t}) per unit mass $\sigmaT/m$. For interactions with velocity dependent differential cross sections, we refer to each model by using the values of $u$ (defined by Equation \ref{eq:vel_dep_normalization}) and $w$ (defined by Equation \ref{eq:vel_dep}) for each model. The self-interaction parameters for all the models explored in this paper are summarized in Table \ref{tab:parameters}.

\begin{table}[t]
    \centering
    \begin{tabular}{|c|c|c|}
    \hline
    \multicolumn{3}{|c|}{Velocity-independent}\\
    \hline
    Type & $\sigma/m$ & $\sigmaT/m$ \\
    \hline
    \hline
         Isotropic & $1\,{\rm cm^2/g}$ & $0.5\,{\rm cm^2/g}$ \\
    \hline
         Isotropic & $2\,{\rm cm^2/g}$ & $1\,{\rm cm^2/g}$ \\
    \hline
         Anisotropic & - & $1\,{\rm cm^2/g}$ \\
    \hline
         Anisotropic & - & $3\,{\rm cm^2/g}$ \\
    \hline
    \end{tabular}
    \qquad
    \begin{tabular}{|c|c|}
    \hline
    \multicolumn{2}{|c|}{Velocity-dependent}\\
    \hline
         w & u \\
    \hline 
    \hline
        $500\,{\rm km/s}$ & $1000\,{\rm km/s}$ \\
    \hline
        $1600\,{\rm km/s}$ & $2000\,{\rm km/s}$ \\
    \hline
        $1000\,{\rm km/s}$ & $2000\,{\rm km/s}$ \\
    \hline
    \end{tabular}
    \caption{Summary of the self-interaction parameters explored in this paper. The angular dependence of the velocity-independent anisotropic cross sections is defined in Equation \ref{eq:anisotropic}. For velocity-dependent cross sections, the parameters $w$ and $u$ are defined by Equations \ref{eq:vel_dep} and \ref{eq:vel_dep_normalization} respectively. The scaling of the cross section with velocity for these models is shown in Fig. \ref{fig:cross_section}. }
    \label{tab:parameters}
\end{table}

%-----------------------------------------%
\subsection{Effects on individual halos: the most massive halo in the simulation}
%-----------------------------------------%

\begin{figure}[t]
\begin{center}
\includegraphics[width=0.9\linewidth]{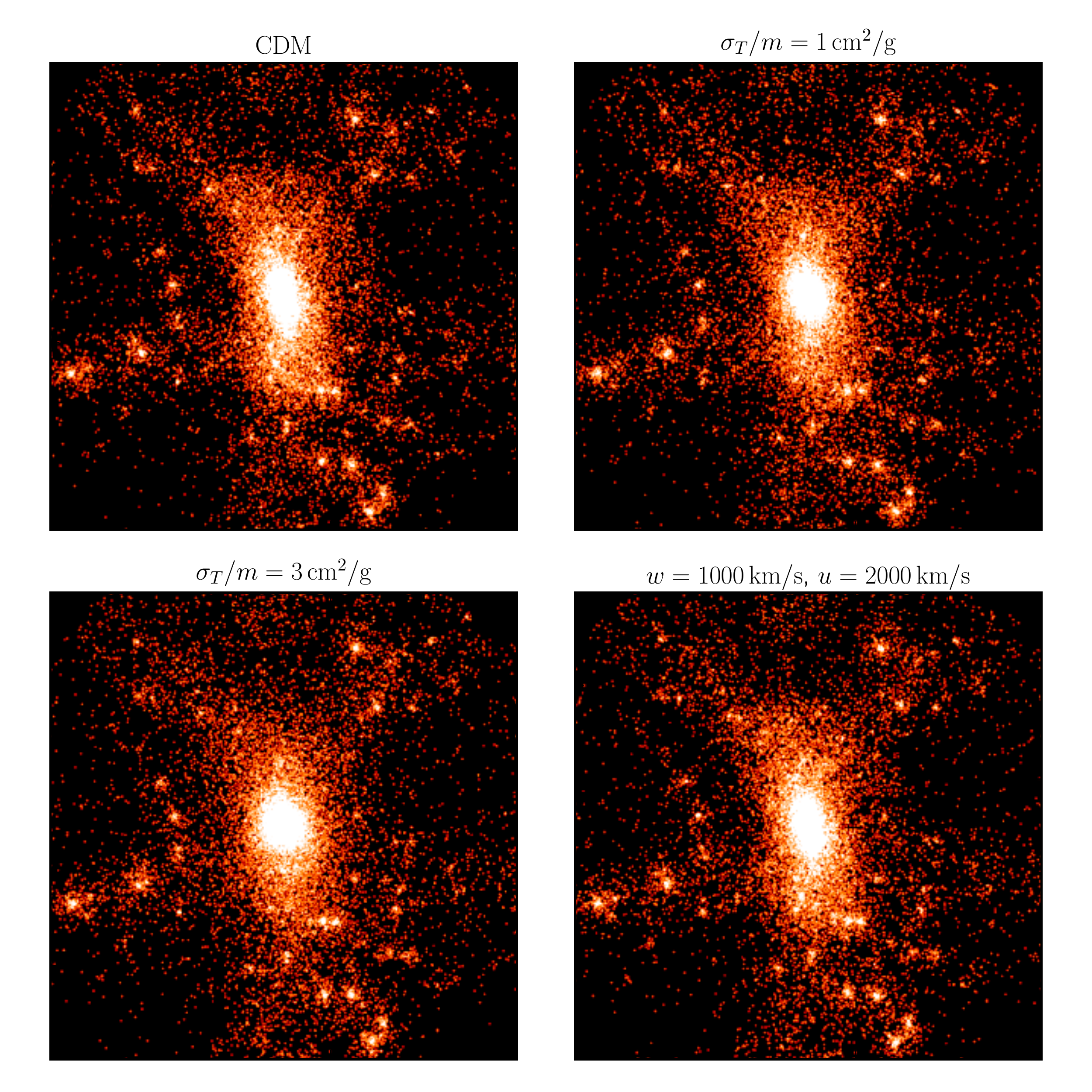}
\end{center}
\caption{Projected distribution of dark matter particles in a $10\,\hMpc$ region around the most massive object in the simulation volume for CDM and different SIDM scenarios. Increasing the strength self-interactions makes the halo rounder, and puffs up the central region.}
\label{fig:scatter_subhalos}
\end{figure}

We begin by studying the effects of self-interactions on the most massive halo in the simulation box. This halo has a virial mass of $4.75\times 10^{15} \hMsun$ with $\sim 6\times 10^5$ particles within the virial radius in the CDM simulation, which allows us to examine its radial mass distribution over a fairly wide range of radii. While this halo is somewhat more massive than the mass range we analyze in the subsequent sections, it is instructive to study the effects of self-interactions on a single well-resolved halo before moving to considering average properties of halos measured from stacked profiles.

First, to illustrate the effect of self-interactions on the overall distribution of dark matter, Fig. \ref{fig:scatter_subhalos} shows 
distribution of mass around the most massive halo projected along the $z$-axis for different self-interaction scenarios. Apart from the CDM simulation, we pick three self-interaction models which best illustrate the differences. The  feature that is most evidently visible is that as the cross-sections for the velocity-independent self-interactions increase, the halos become significantly rounder. This is true for both isotropic and anisotropic differential cross sections. This effect has already been seen for the isotropic cross sections in previous studies \cite{Moore:2000,Peter:2012jh}. On the other hand, we find that the shape of the mass distribution in this halo stays almost the same  in the case of the velocity-dependent cross-section. This is because for our choice of the parameters of velocity-dependent interactions, the effective cross sections are quite small at the typical relative velocities within the halo. Specifically, most interactions occur at relative velocities deep in the $v^{-4}$ scaling region shown in Fig. \ref{fig:cross_section}.

\begin{figure}
\begin{center}
\includegraphics[width=0.95\linewidth]{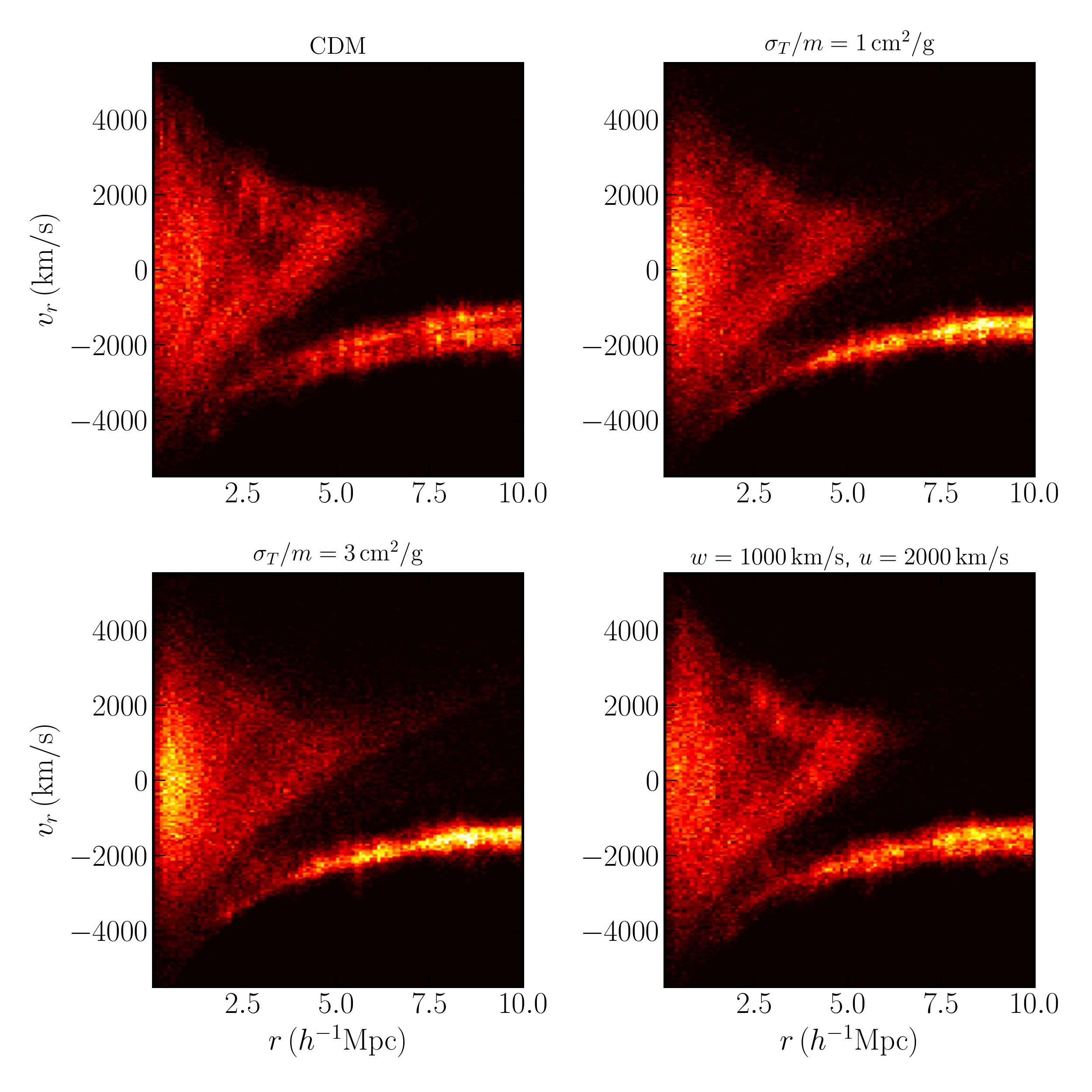}
\end{center}
\caption{Radial phase space diagram of the most massive halo in the simulation in the different models of self-interaction. All panels use the same color stretch.}
\label{fig:phase_space_1e15}
\end{figure}

\begin{figure}
\begin{center}
\includegraphics[width=0.95\linewidth]{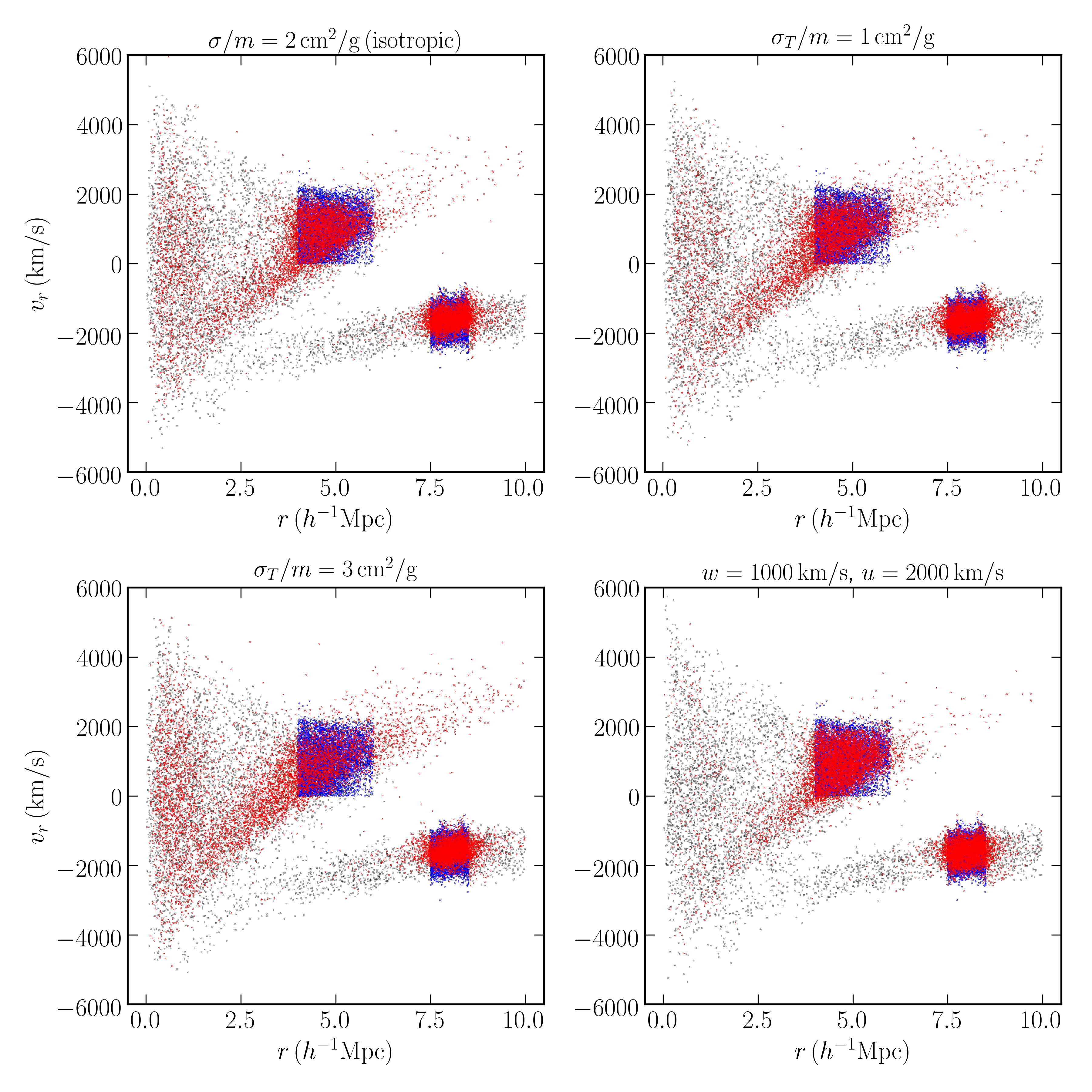}

\end{center}
\caption{The phase space location of corresponding particles in CDM and SIDM. Two sets of particles are chosen from the CDM sims (blue dots), one set corresponds to particles near splashback radius with $3\hMpc < r < 4\hMpc$  and $v_r > 0$ km/s and the other set corresponds to infall particles, $7.5 \hMpc< r < 8.5\hMpc$. The red points correspond to particles in the SIDM simulations with the same particle IDs as the particles from CDM. The black scatter plot in the background shows the CDM particles to guide the eye.}
\label{fig:corres}
\end{figure}

Next, we investigate the phase space distribution of the particles in the halo in Fig. \ref{fig:phase_space_1e15}. Instead of examining the 6-dimensional phase space, we focus on the 
reduced two-dimensional $r-v_{r}$ plane, where $v_r$ is the radial component of the relative velocity between each particle and its respective halo.
There are several differences to be noted when comparing the SIDM panels to the CDM panel, as well as between the different SIDM panels. First, deep inside the multi-stream region of the halo, individual streams remain visibly distinct in the CDM case. On the other hand, they appear more dispersed in all of the SIDM halos due to interactions with other streams. Second, we see two distinct infall streams in the CDM simulation, clearly visible at $\sim 5-10\,\hMpc$. We find that these streams correspond to particles falling from different directions, with different velocity profiles due to the potential being slightly anisotropic in the case of CDM. However, when we consider the upper right and lower left panels of Fig. \ref{fig:phase_space_1e15}, we see that the presence of two streams is much less obvious, pointing to the potential becoming more isotropic in the presence of self-interactions. Note that we do not see this effect as distinctly in the bottom right panel. This is the case of the velocity-dependent interactions which is consistent with the halo shape remaining unchanged for these models at the cross-sections we have investigated in this paper. 

It is also instructive to understand how the self-interactions change the positions and velocities of individual particles in different parts of the halo. Fig. \ref{fig:corres} illustrates the effects of self-interactions on individual particles in various parts of phase space around the most massive object in the box. The black points in each panel are particles from the CDM simulation and are used to guide the eye. The blue points in each panel are also those from the CDM simulation, but from two specific regions of phase space: one set of particles from the infall stream, and the other set of particles moving out towards splashback. The red points denote the phase space positions of the same particles (matched by the particle IDs) from the various SIDM simulations. We note that for the infall stream, the extra interactions have some effect on the positions of the particles but do not change them too much - in particular, none of the particles from the infall stream becomes unbound. On the other hand, for the set of particles near splashback, the interactions can lead to some particles becoming unbound, as is seen by the presence of particles with positive radial velocities beyond $\sim 5\,\hMpc$. Further we note that a fraction of the particles in the SIDM simulations have turned around before reaching the CDM splashback. These are particles that lost energy due to self-interactions, thereby putting them on more tightly bound orbits, with smaller orbit times.

%-------------------------------------------------------------------
\subsection{Effect of angular dependence of the differential cross section on stacked profiles}

%--------------------------------------------------------
\label{sec:drag}

\begin{figure}
\begin{center}
\includegraphics[width=0.45\linewidth]{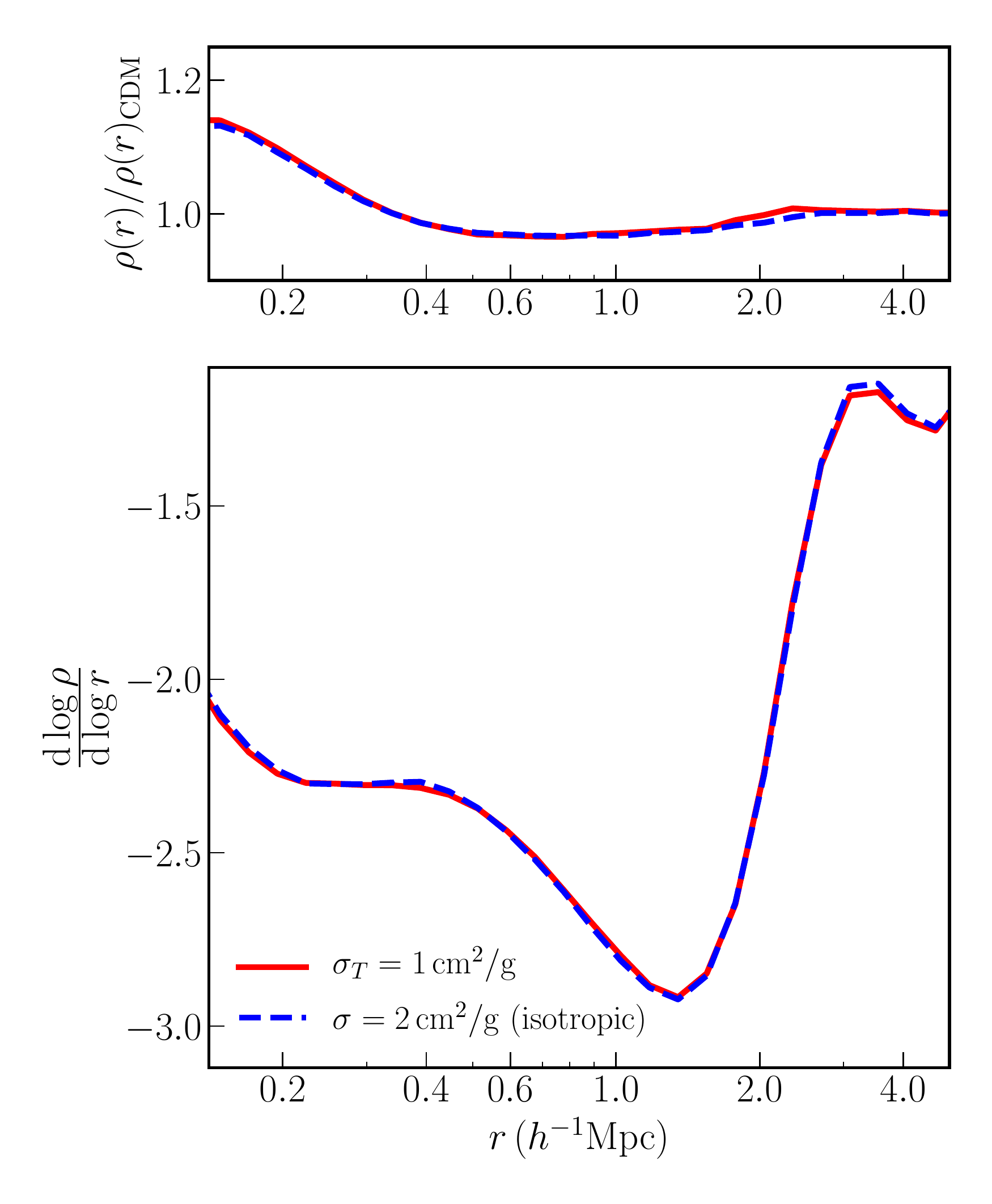}
\includegraphics[width=0.45\linewidth
]{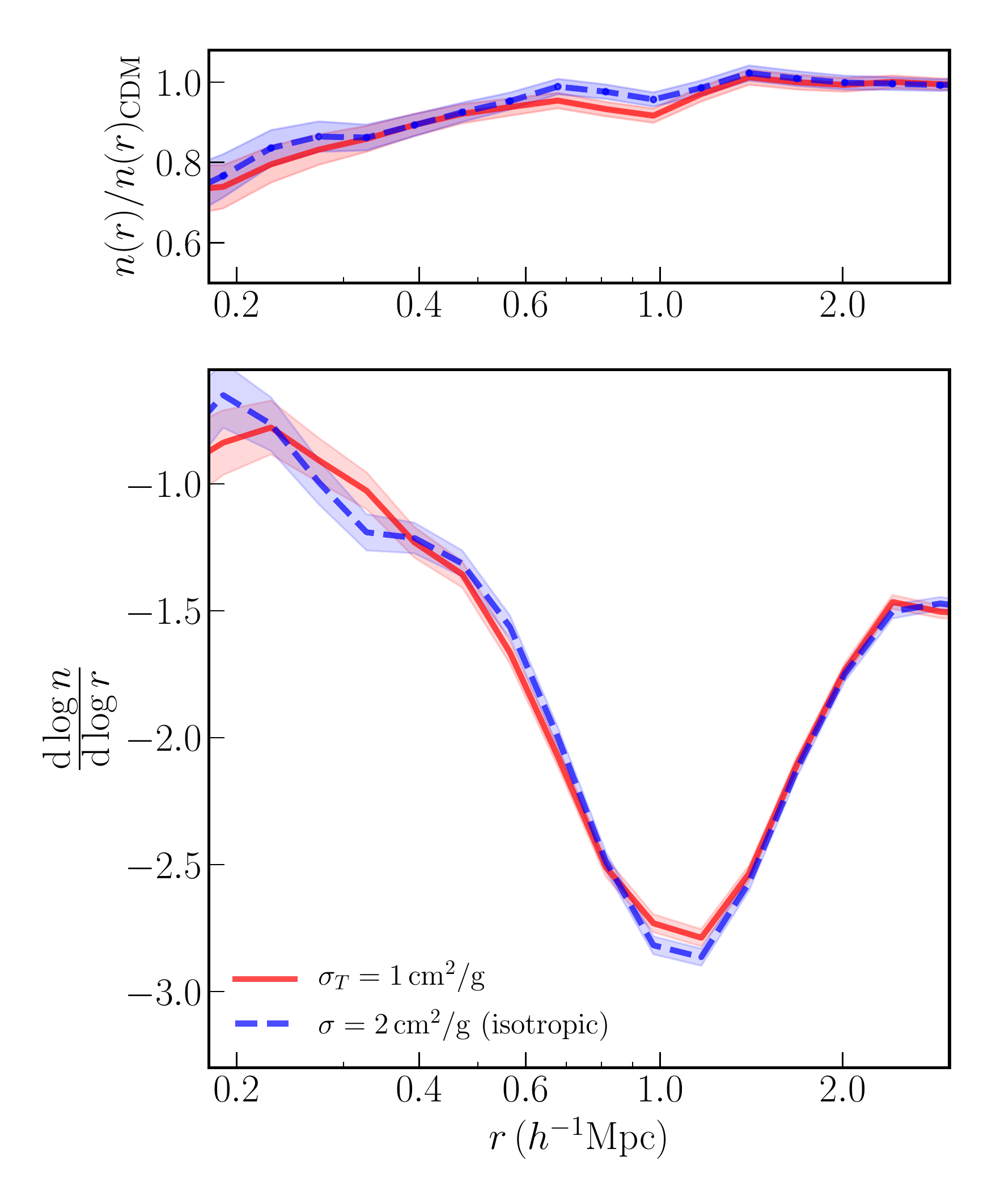}
\end{center}
\caption{Comparison of profiles and slopes around clusters in the mass range $1-2\times10^{14}\hMsun$ in matched $\sigmaT$ simulations with velocity-independent isotropic and anisotropic differential cross sections. Top left: Ratio of the stacked density profiles normalized by the CDM profile. Bottom left: Logarithmic slope of the particle density profile. Top right: Ratio of the stacked subhalo counts around clusters in the chosen mass range. Bottom right: Logarithmic slope of the stacked number count profile.}
\label{fig:drag_appendix}
\end{figure}

For the velocity independent anisotropic cross sections given by Equation \ref{eq:anisotropic}, small angle scatterings with low momentum exchange dominate over scattering events with large scattering angles and momentum exchange. On the other hand, for isotropic cross sections, the scattering angle is distributed uniformly in $\cos\theta$. \cite{Kahlhoefer:2013dca} suggested that the angular dependence of the cross section could lead to qualitatively different effects as compared to isotropic cross sections - specifically a coherent ``drag" on particles accreting on to massive clusters. Since the splashback radius of clusters is set by the apocenters of particles accreting onto them, the presence of an extra force could shrink the orbits and therefore, shrink the splashback radius \cite{More:2016vgs}.

For the rest of the paper, we will focus on the effect of self-interactions on the stacked particle and subhalo profiles, as well as the splashback radius, of clusters in the mass range $1\times 10^{14}\hMsun$ to $2\times 10^{14}\hMsun$, and so we first investigate the effect of the angular dependence of the differential cross-section on these specific observables. To carry out this study, we need to match some macroscopic property of the two types of interactions, and for simplicity, we restrict ourselves to velocity-independent cross sections. As pointed out in Section \ref{sec:anisotropic}, the total cross section diverges for the angular dependence considered here, and therefore is not a useful quantity to hold fixed for the comparison. A more natural choice is the momentum transfer cross section $\sigmaT$, which is well defined for both the isotropic as well as the anistropic differential cross sections. We will therefore compare simulations of isotropic and anisotropic cross sections for which $\sigmaT$ has been matched. In particular, Equation \ref{eq:sigma_t} can be used to show that, for isotropic cross sections, $\sigma/m=2\,{\rm cm^2/g}$ implies $\sigmaT/m=1\,{\rm cm^2/g}$. 

In Fig. \ref{fig:drag_appendix}, we present the comparison of the results from the two simulations. The top left panel shows the ratio of the stacked particle density  profiles around clusters in the chosen mass range from each self-interaction scenario normalized by the profile from the CDM simulation. The bottom left panel shows the logarithmic slope of the stacked particle profile. Both panels illustrate that once $\sigmaT$ has been matched, there are no clear signals of actual angular dependence of the cross sections on the profile or its slope. Note that the splashback radius of the stacked sample, as defined by the minimum of the logarithmic slope, shows no movement between the two forms of self-interactions.

We also investigate the effects of the angular dependence on the number counts of subhalos around these clusters. This is presented on the right hand panels of Fig. \ref{fig:drag_appendix}. The top right panel again plots the ratio of the number counts from each simulation normalized by the number counts from the CDM simulation. The bottom right panel plots the logarithmic derivative of the number count profile. Comparing the red and blue curves in each panel, we find no statistically significant difference between the results. We therefore conclude that the angular dependence of the self-interaction cross section does not play a significant role in the stacked density profile or satellite counts around clusters, for interactions of the form given by Eqn.\ \eqref{eq:anisotropic}.  This also tells us that we can interpret the results shown in the subsequent sections by keeping in mind that isotropic interactions with $\sigma/m=1\,{\rm cm^2/g}$ gives equivalent effects to anisotropic interactions with $\sigmaT/m=0.5\,{\rm cm^2/g}$; and  anisotropic interactions with $\sigmaT/m=3\,{\rm cm^2/g}$ correspond to isotropic interactions with $\sigma/m=6\,{\rm cm^2/g}$.

%-----------------------------------------------------------------------
\subsection{Mean Density profile and splashback radius of dark matter halos}
%-----------------------------------------------------------------------

\begin{figure}
\begin{center}
\includegraphics[width=0.45\linewidth, trim= 0.4in 0in 0in 0in]{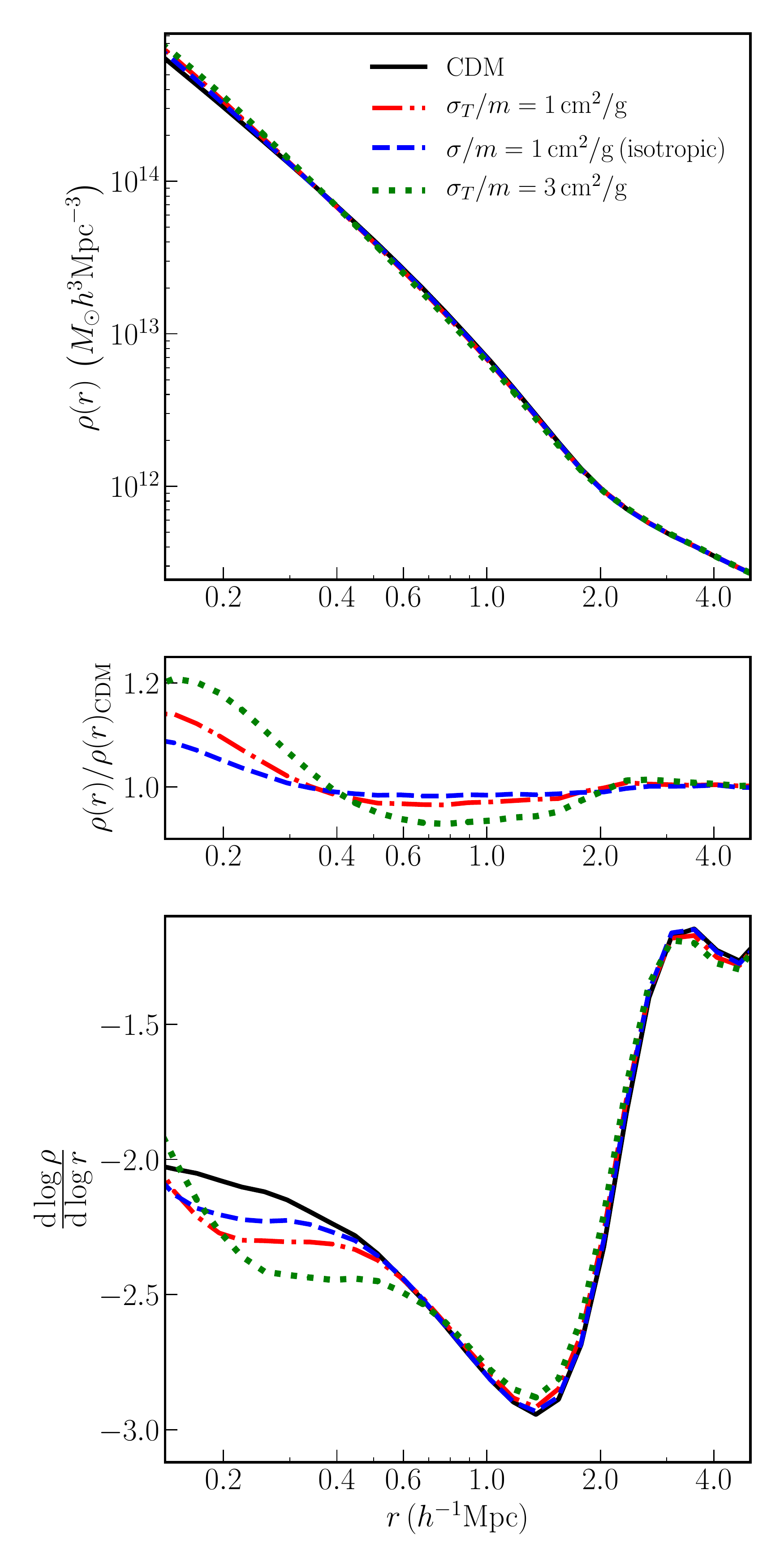}
\includegraphics[width=0.45\linewidth,trim= 0.4in 0in 0in 0in]{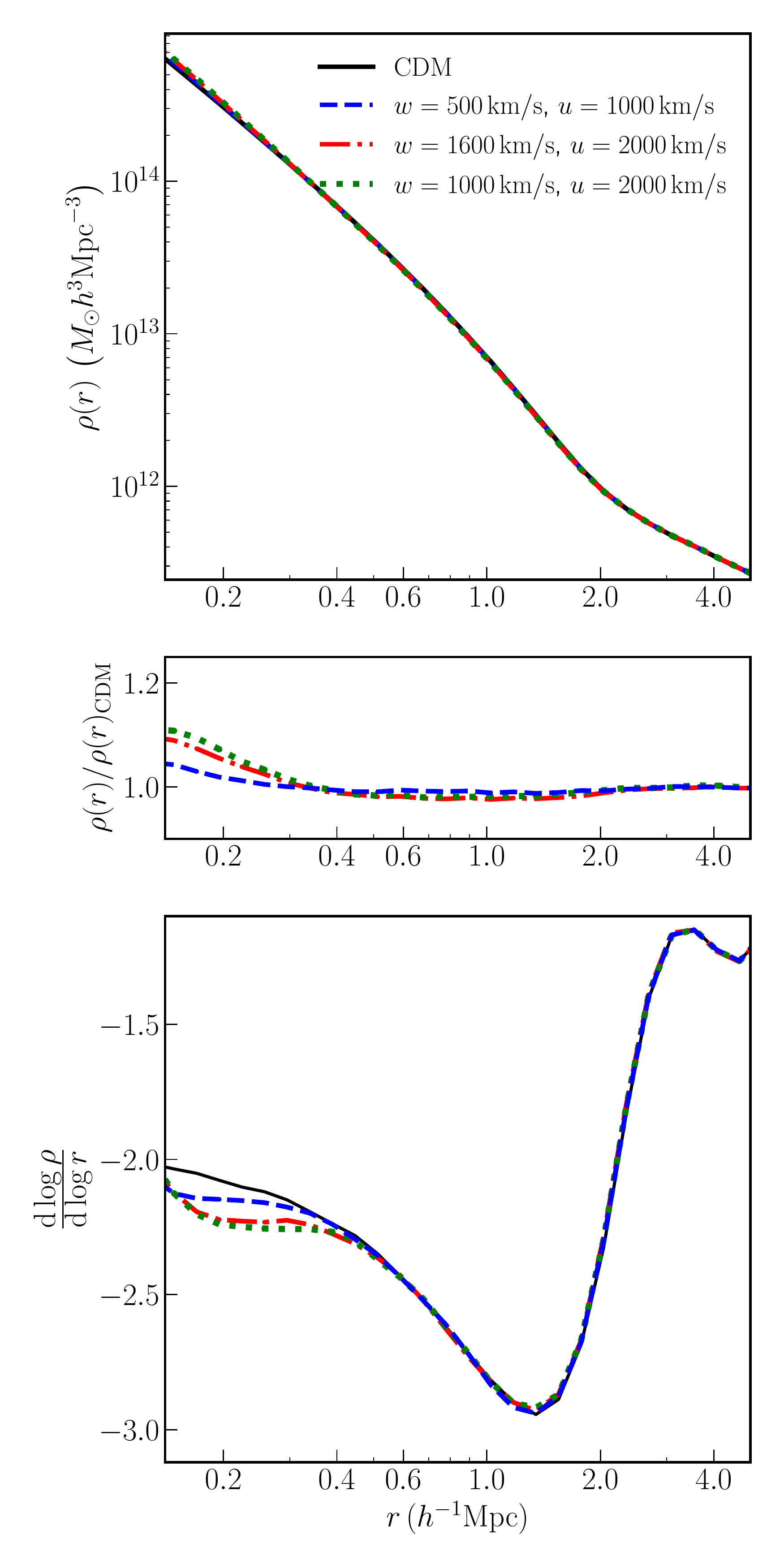}
\end{center}
\caption{Top: Stacked density profile of particles around halos in the mass range $1-2\times 10^{14}\hMsun$ in the velocity-independent (left panels) and velocity-dependent (right panels) SIDM scenarios. Middle: Ratio of the stacked density profile in each self-interaction scenario to the profile in the CDM case. Bottom: Logarithmic derivative of the slope of the density profile. Self-interactions actually steepen the density profiles compared to CDM on scales just larger than the scale radius ($\sim 0.14 \hMpc$). The position of splashback does not depend sensitively on the strength of the self-interactions.}
\label{fig:halo_profile_particles}
\end{figure}

In  this section we focus on massive cluster halos with virial masses in the range  $1-2\times 10^{14} \hMsun$. There are $\sim 20000$ halos in this mass range in our $1 \hGpc$ box. We first examine the stacked dark matter density profiles for different models of self-interaction. While the effect of self-interactions has been well-studied with respect to changes of the profile deep inside the virial radius, here we focus on the effects of self-interactions on the profile at radii around  the splashback radius, i.e. scales larger than the typical scale radius of these halos.

The top panel of Fig. \ref{fig:halo_profile_particles} shows the stacked mean density profiles in both velocity independent and velocity dependent simulations. The middle panel shows the ratio of density profiles as a function of radius. Depending on the cross-section, in the velocity independent case there can be about $20\%$ increase in the densities in SIDM halos before it flattens further inside the halos. Overall, we can see that concentration of dark matter profile increases with increasing $\sigmaT/m$. The results for anisotropic and isotropic cross-sections are qualitatively similar (recall that the isotropic cross section with $\sigma/m=1\,{\rm cm^2/g}$ corresponds to $\sigmaT/m\approx 0.5\,{\rm cm^2/g}$). 

\begin{figure}
\begin{center}
\includegraphics[width=0.45\linewidth]{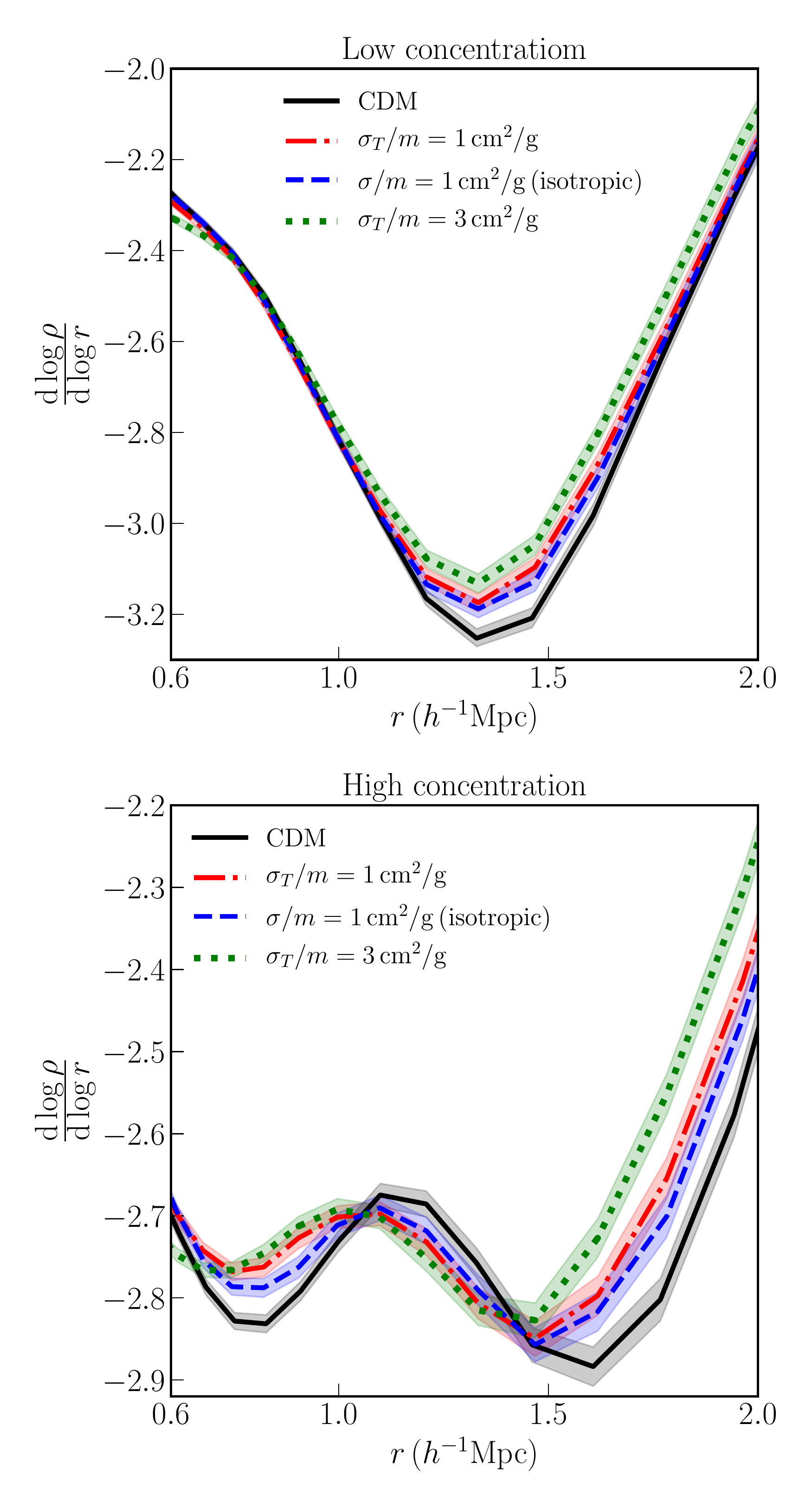}
\includegraphics[width=0.45\linewidth]{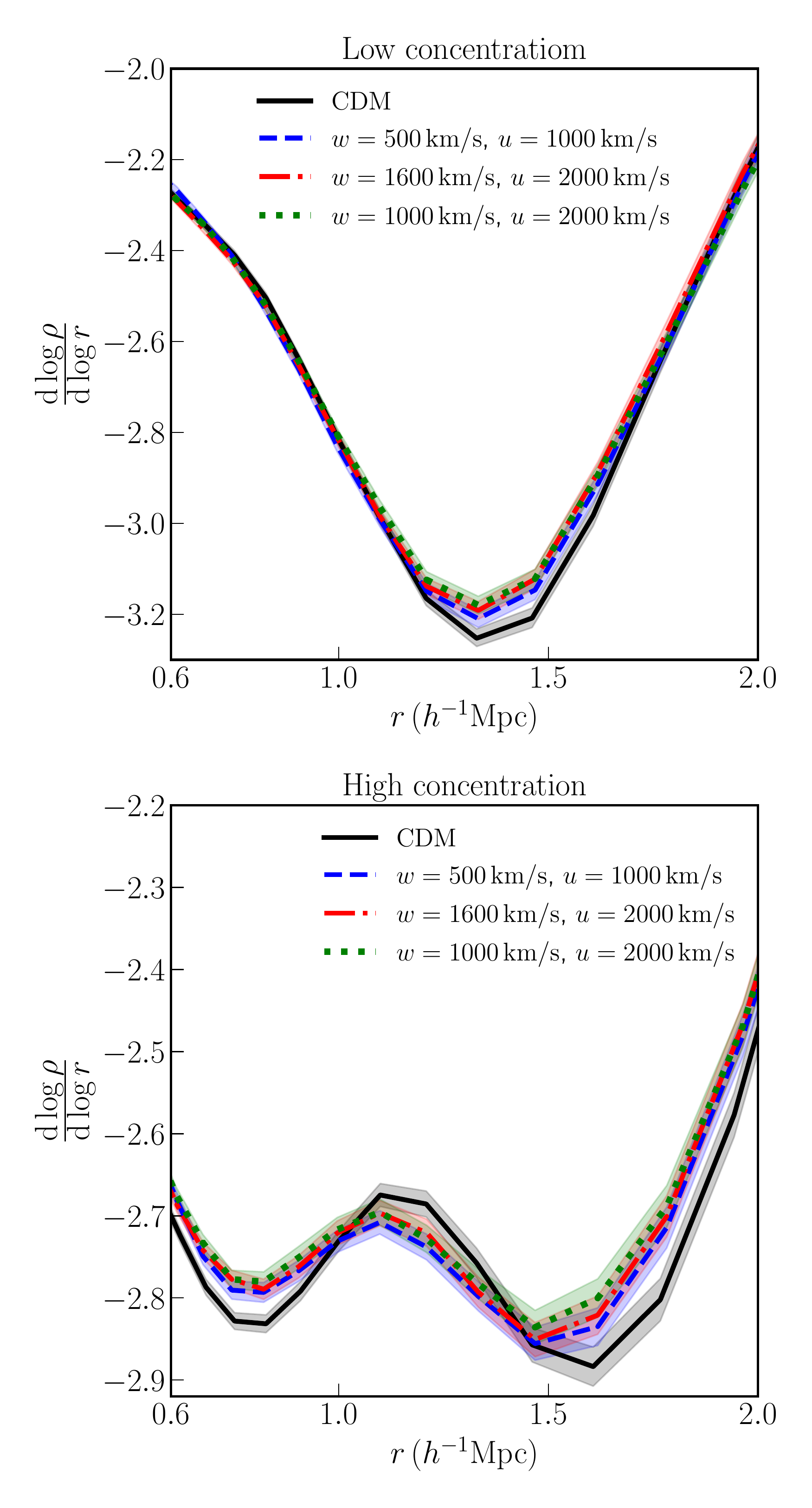}
\end{center}
\caption{Stacked logarithmic derivative of the particle density profile of clusters in the mass range $1-2\times10^{14}\hMsun$ for halos above median concentration (top panel) and for halos below median concentration (bottom panel). Left panels show the results for velocity-independent self-interactions, while the right panels are for the velocity-dependent self-interactions. The splashback radius does not change for low concentration halos in the presence of self-interactions, but the location of splashback moves inward for the high concentration clusters.}
\label{fig:conc_splashback}
\end{figure}

For the velocity-dependent interactions, we find that the models with $w=1600\,{\rm km/s}$, $u=2000\,{\rm km/s}$ and $w=1000\,{\rm km/s}$, $u=2000\,{\rm km/s}$ produce roughly similar effects on the stacked density profiles, even though they have very different cross sections at low velocities as can be seen in Fig. \ref{fig:cross_section}. The reason for this is that the models have similar cross section at the typical relative velocities between particles of the clusters in our sample ($\sim 1500\,{\rm km/s}$). On the other hand, the effect of the model with $w=500\,{\rm km/s}$, $u=1000\,{\rm km/s}$ is much smaller, even though at low velocities the cross section is the same as that of the model with $w=1000\,{\rm km/s}$, $u=2000\,{\rm km/s}$. Once again, this is because at the typical collision velocities relevant to the chosen clusters, the former has much lower cross sections.

In the bottom panels of Fig. \ref{fig:halo_profile_particles}, we plot the logarithmic derivative of the stacked density profiles of massive halos. The edge of the multi-streaming region of the halos, as defined by the splashback radius, is traced by the minimum of the slope profile. The figure shows that effect of self-interactions on the location of the splashback radius is not visible when all halos are stacked.  This is true even when the cross section is increased to $\sigmaT/m = 3\, \mathrm{cm^2/g}$ (for anisotropic interactions), which corresponds to a regime of self-interactions already ruled out by the constraints derived from observations of the Bullet Cluster \cite{Markevitch:2003at,Robertson:2016xjh}. 
The only effect of self interactions on the stacked profile appears to be a mild shallowing of the splashback feature, instead of a coherent movement inward. 

The mean profile discussed so far is an average over many different halo histories. In the CDM paradigm, cluster mass objects that form early tend to have more concentrated profiles than younger or late forming halos \cite{Wechsler:2001cs}. Since the interaction rate in SIDM models is proportional to the local density, it is expected that for older halos (and more concentrated halos), the number of interactions  near the centre during a particle's orbit should be larger than in their younger counterparts. To test this, we match halos in the SIDM simulations to their CDM counterparts and look at the average behaviour of the halo profiles in populations that have different CDM concentrations. 

Fig. \ref{fig:conc_splashback} shows the splashback region of the logarithmic-derivative profile for different models of SIDM.  The top and bottom panel represent the derivatives from the average profiles of halos with concentrations that are lower and higher than the median concentration, respectively. The left panels show the comparison between CDM and velocity-independent SIDM models, while the right panels show the comparison between CDM and the velocity-dependent SIDM models. The location of the splashback radius remains unchanged in halos with low concentration for all SIDM models. On the other hand, the lower panels clearly show that SIDM halos with high concentration CDM counterparts indeed have a smaller splashback radius compared to CDM. As expected, larger interaction cross-sections produce larger shifts in the position of splashback in the high-concentration halos. This behavior is exhibited by both the velocity-dependent and velocity-independent scenarios, and the shift is roughly $20\%$  for a realistic cross section of $\sigmaT=1\, \mathrm{cm^2/g}$.

%------------------------------------
\subsection{Distribution of subhalos}
%------------------------------------

\begin{figure}
\begin{center}
\includegraphics[width=0.45\linewidth]{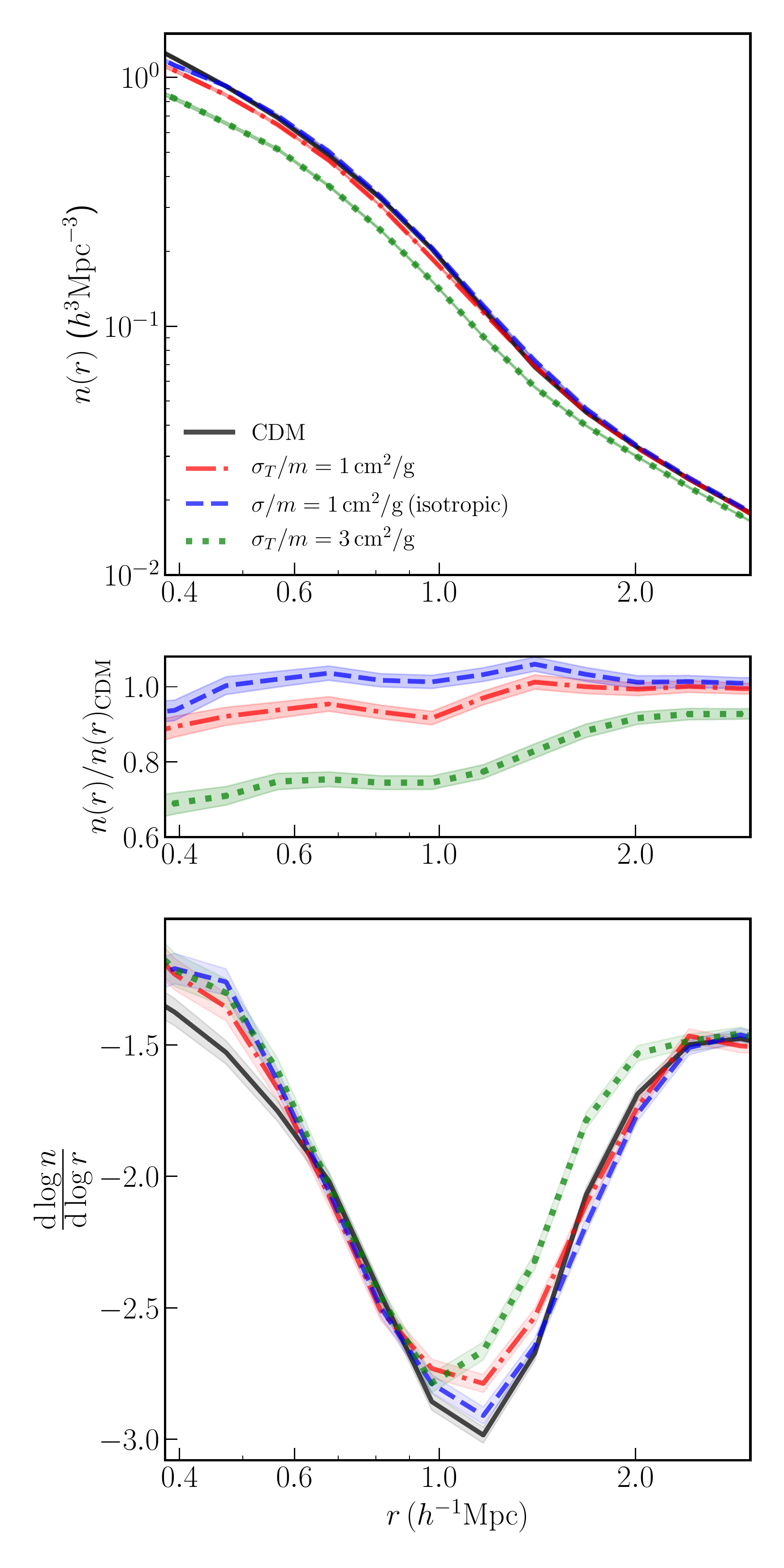}
\includegraphics[width=0.45\linewidth]{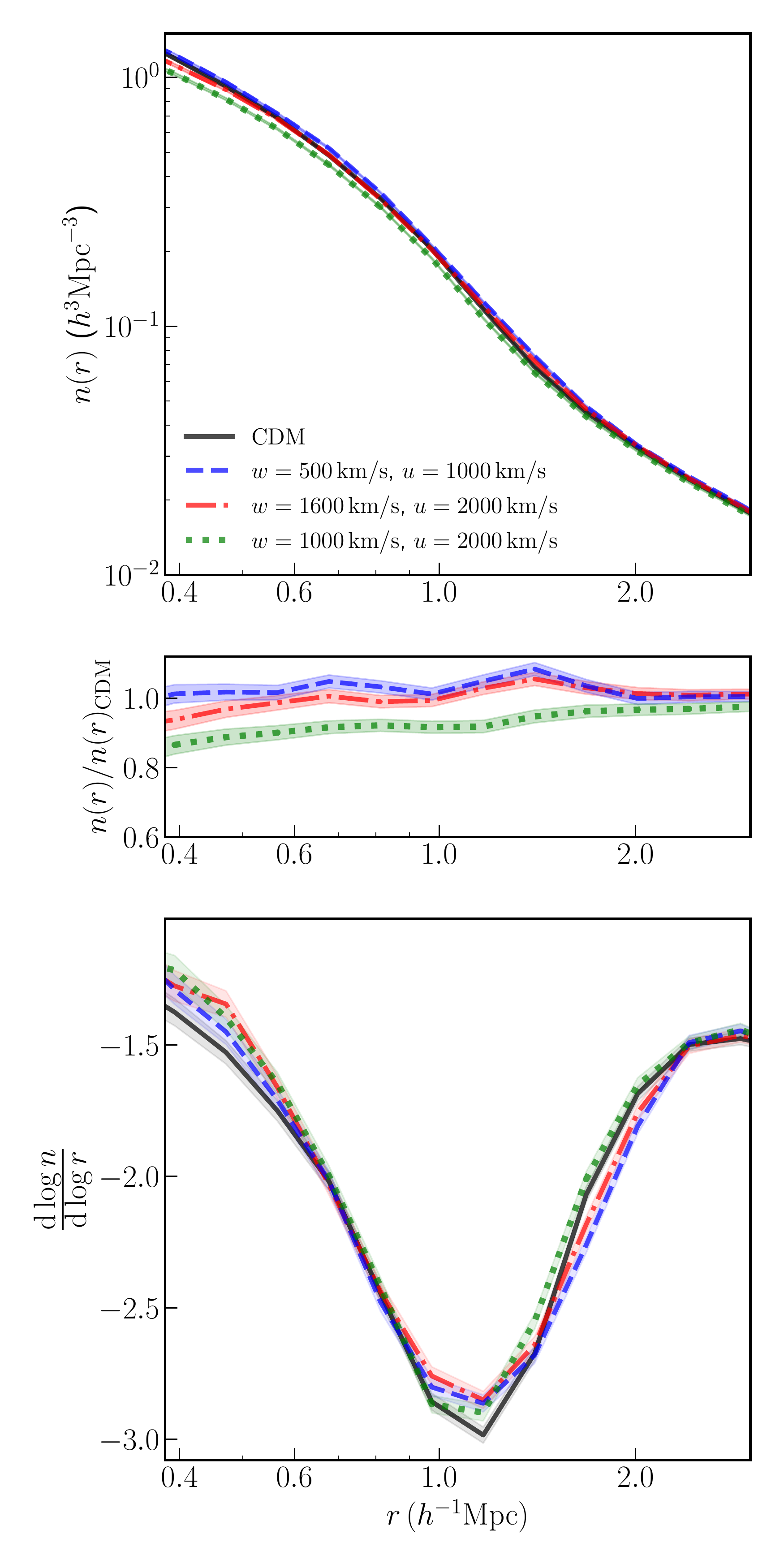}
\end{center}
\caption{Top: Stacked number density profiles of subhalos around the same halos as in Fig. \ref{fig:halo_profile_particles} in the velocity dependent and velocity-independent SIDM scenarios. Middle: Ratio of the number counts of subhalos for each self interacting scenario to the CDM number counts. Bottom: Logarithmic derivative of the slope of the subhalo number profiles. Presence of self-interactions reduces the number counts of subhalos at all scales. The position of splashback is not very sensitive to SIDM, but becomes shallower as the cross sections increase. Left panels show results for velocity-independent interactions, while the right panels show results for the velocity-dependent interactions. Shaded regions represent the statistical uncertatinties on each quantity. }
\label{fig:halo_profile_subhalos}
\end{figure}

In the CDM paradigm, low-mass halos form first, and then merge into larger structures. Some of these original small structures survive in the larger halos as subhalos, which host satellite galaxies in galaxy clusters. It is therefore interesting to study the distribution of substructure within halos and how it is affected by self-interactions among dark matter particles. Dynamics of subhalos should generally be similar to that of dark matter particles, except that subhalos can be disrupted due to tidal forces of their host halo and may experience dynamical friction (significant for subhalos with masses $M_{\rm sub}\gtrsim 0.01 M_{\rm host}$). In addition, in the presence of self-interactions of dark matter particles there can be extra loss of mass from the  subhalo, as they fall into the virial structure of the host with high velocities. This evaporation process \citep{GnedinOstriker:2001} should affect their radial distributions within the host. The self-interactions can also produce a ``drag'' on the subhalos, which means that subhalos on more radial orbits will tend to show smaller apocenters than in the CDM model without self-interactions. 

The top panel of Fig. \ref{fig:halo_profile_subhalos} shows the stacked number density profiles of subhalos around host halos of mass $1-2\times 10^{14} \hMsun$. The left and right panels correspond to velocity-independent and velocity-dependent cross-sections, respectively. The middle panels show the ratio of the number density of subhalos in each self-interacting scenario to the number count in the CDM simulation. The bottom panel shows the logarithmic derivative of the slope of the number density profiles around these halos. Given our resolution limits, only we only use subhalos for which $V_{\rm peak}>200\, {\rm km/s}$ in this analysis. Since the typical peak masses of the these subhalos are  $\gtrsim 1\%$  of the mass of their host, they are expected to experience dynamical friction. The dynamical friction manifests itself in the fact that the splashback radius that we find for the subhalos is smaller than the splashback radius as found using the mass profiles in Fig. \ref{fig:halo_profile_particles}. 

Fig. \ref{fig:halo_profile_subhalos} shows that the abundance of subhalos is suppressed within the halo in the velocity-independent models as compared to the CDM simulation - for anisotropic cross section with $\sigmaT/m=3\,{\rm cm^2/g}$ showing suppression of $\sim 30\%$ at $r\sim 0.5\hMpc$. In general, the splashback feature, shown in the bottom panel, becomes shallower as the cross-section is increased. At these cross-sections subhalos experience significant self-interactions all throughout the host, it should also be noted that the tidal forces on these subhalos are enhanced as well due to the relatively steeper particle profile within the host (Fig. \ref{fig:halo_profile_particles} bottom, left panel). Further, subhalos that are on eccentric orbits and reach the central regions where the density is higher, and undergo more collisions per unit time, are disrupted more frequently than subhalos that have large pericenters. These effects add up to make the splashback feature shallower than for the CDM case, since all the subhalos that make it back to the splashback radius in the CDM case are not able to so in the presence of self-interactions. We note that  when the cross section is turned up to very high values, i.e. anisotropic scattering with $\sigmaT/m=3\,{\rm cm^2/g}$, the splashback position also exhibits a shift towards smaller radius, and the magnitude of this shift is close to $20\%$ of the position of the CDM splashback.

\begin{figure}
\begin{center}
\includegraphics[width=1.0\linewidth]{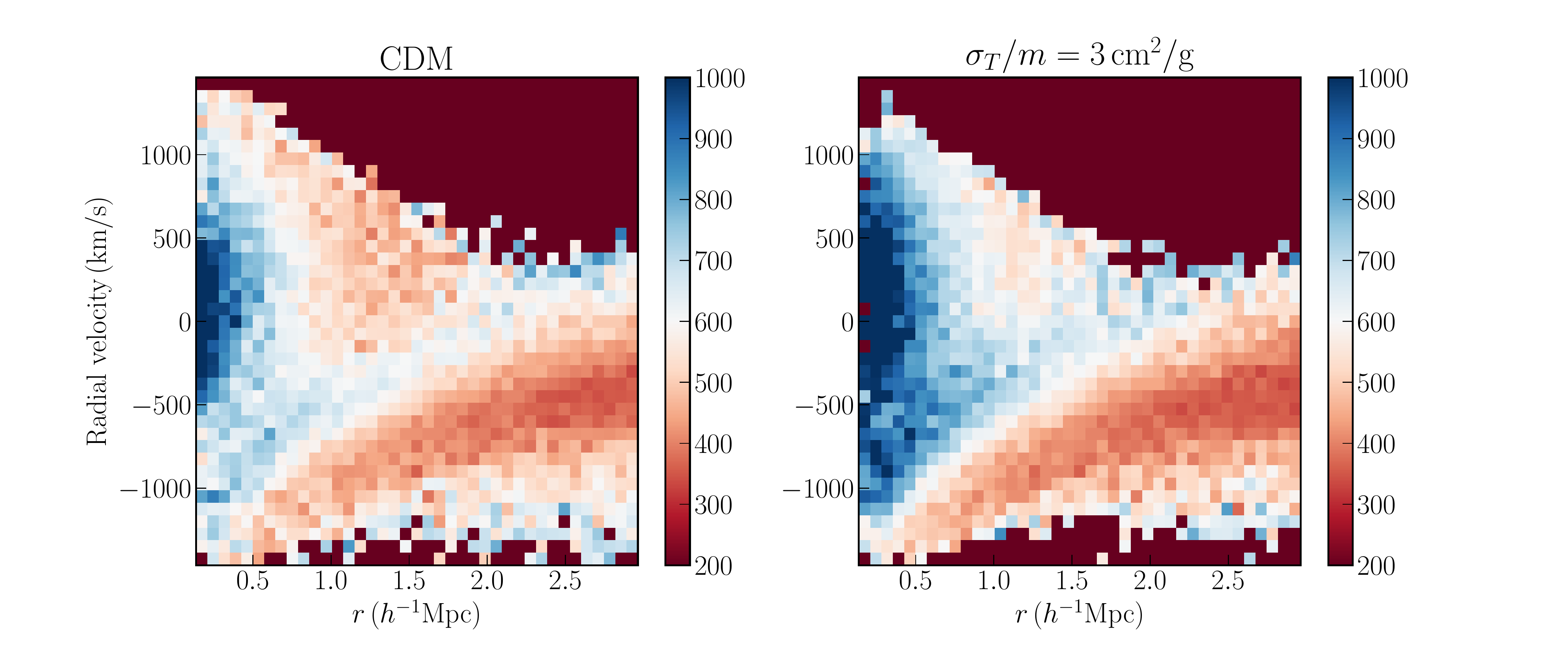}
\end{center}
\caption{Stacked radial phase space plot of subhalos around clusters in the mass range $1-2\times10^{14}\hMsun$. Each pixel is colored by the value of the mean tangenital velocity of all subhalos corresponding to that pixel. We only populate pixels with more that 5 subhalos to make the phase space envelope more clearly visible. For CDM (left panel), we find that the subhalo propulation near splashback is dominated by those that have low tangential velocities i.e. their orbits were mostly radial. For the strongest SIDM model (right panel) we find that these subhalos with radial orbits have been preferentially destroyed due to closer pericenters to the host's center. The lack of these subhalos leads to lower measured splashback radius for the subhalos in the model with $\sigmaT/m=3\,{\rm cm^2/g}$ (bottom left panel of Fig. \ref{fig:halo_profile_subhalos}).}
\label{fig:vtan_subhalos}
\end{figure}

To understand this shift, we consider the stacked radial phase space structure around these clusters as sampled by the subhalos used in our analysis. However, we also retain information about the tangential velocities of the subhalos in different parts of the phase space. This is shown in Fig. \ref{fig:vtan_subhalos}, where each pixel in the radial phase space has been colored by the mean tangential velocity of all subhalos that belong to the pixel. We only populate pixels with more that 5 subhalos to make the phase space envelope more clearly visible. For CDM, shown on the left panel of Fig. \ref{fig:vtan_subhalos}, we find that near splashback, the phase space is dominated by subhalos with low tangential velocities. We infer that, for CDM clusters, the subhalos with the largest apocenters, and therefore those that define the splashback radius, are those on preferentially radial orbits. We note that radial orbits are characterized by low pericenters; these orbits extend into the densest parts of the host halo. On the right panel of Fig. \ref{fig:vtan_subhalos}, we make a similar plot for SIDM with $\sigmaT/m=3\,{\rm cm^2/g}$. We find a smaller abundance of subhalos with low tangential velocities near splashback in the presence of strong self-interactions. This is because these subhalos on orbits with low pericenters were destroyed by a combined effect of gravitational tides from the host and self-interaction effects. The self-interactions effects scale with density and relative velocity, and therefore are the strongest for subhalos that reach the densest parts of the host cluster. Since self-interactions preferentially disrupt exactly those subhalos which would otherwise have had the largest apocenters, this shows up in the stacked density profile as a shrinking of the splashback radius. Note that a large fraction of the subhalos with radial orbits need to be disrupted on their very first passage through the halo center for this effect to be visible. This is why the effect discussed above shows up only for the strongest models of self-interactions. 

For the velocity-dependent SIDM simulations we find that the location of the splashback radius for the cross-sections considered in this paper does not move from the CDM case. 
The number density profiles of subhalos in the velocity-dependent simulations are however significantly different from CDM for the case with  $w=1000 \,{\rm km/s}$, and $u=2000 \,{\rm km/s}$. The differences are as large as $\sim 10\%$ out to virial radius. Referring to Fig. \ref{fig:cross_section}, we note that there are two relevant velocity scales for subhalos within clusters - one is the velocity dispersion of particles within subhalos before the subhalo falls into the host halo, and the second is the relative velocity between the subhalos and the host halos. While all three curves in Fig. \ref{fig:cross_section} have $\sigmaT /m\lesssim 1\,{\rm cm^2/g}$ at the velocity scale relevant for the host halo, the blue and green curves have higher interaction cross-sections at the scale of the subhalos' own internal velocity dispersion. The latter makes the subhalos more cored or less bound  even before they fall into the cluster and therefore more likely to be disrupted due to both SIDM interactions or tidal forces compared to the cross-sections corresponding to the red curve. This explains why the green curve in the right panel of Fig. \ref{fig:halo_profile_subhalos} is the most suppressed - the subhalos in that simulation are more likely to be tidally disrupted due to being loosely bound and also have more SIDM interactions within the halo. In comparison, the blue curve which has similar cross-section at low velocities but much lower cross-section at the velocity scales relevant for massive clusters, does not show the same suppression in the number density. Also note that while the red (dot-dashed) and green (dotted) curves from the right hand panel of Fig. \ref{fig:halo_profile_particles} follow each other closely since they have similar cross sections at the cluster velocity scale, the same models produce very different subhalo number counts in the right hand panel of Fig. \ref{fig:halo_profile_subhalos}, as a result of having different cross sections at the velocities relevant for objects of mass $\sim 10^{12}\hMsun$.

We therefore infer that looking only at the suppression of subhalo number-density profiles can lead to degeneracies in the model predictions of velocity dependent and independent cross-sections. However, we can break this degeneracy by combining with measurement of particle density profiles through weak lensing. The particle density profiles are sensitive to the cross-section at the high velocity scale corresponding to the internal dispersion of the cluster ($\sim 1000\, \rm{km/s}$). 

\begin{figure}
\begin{center}
\includegraphics[width=0.45\linewidth, ]{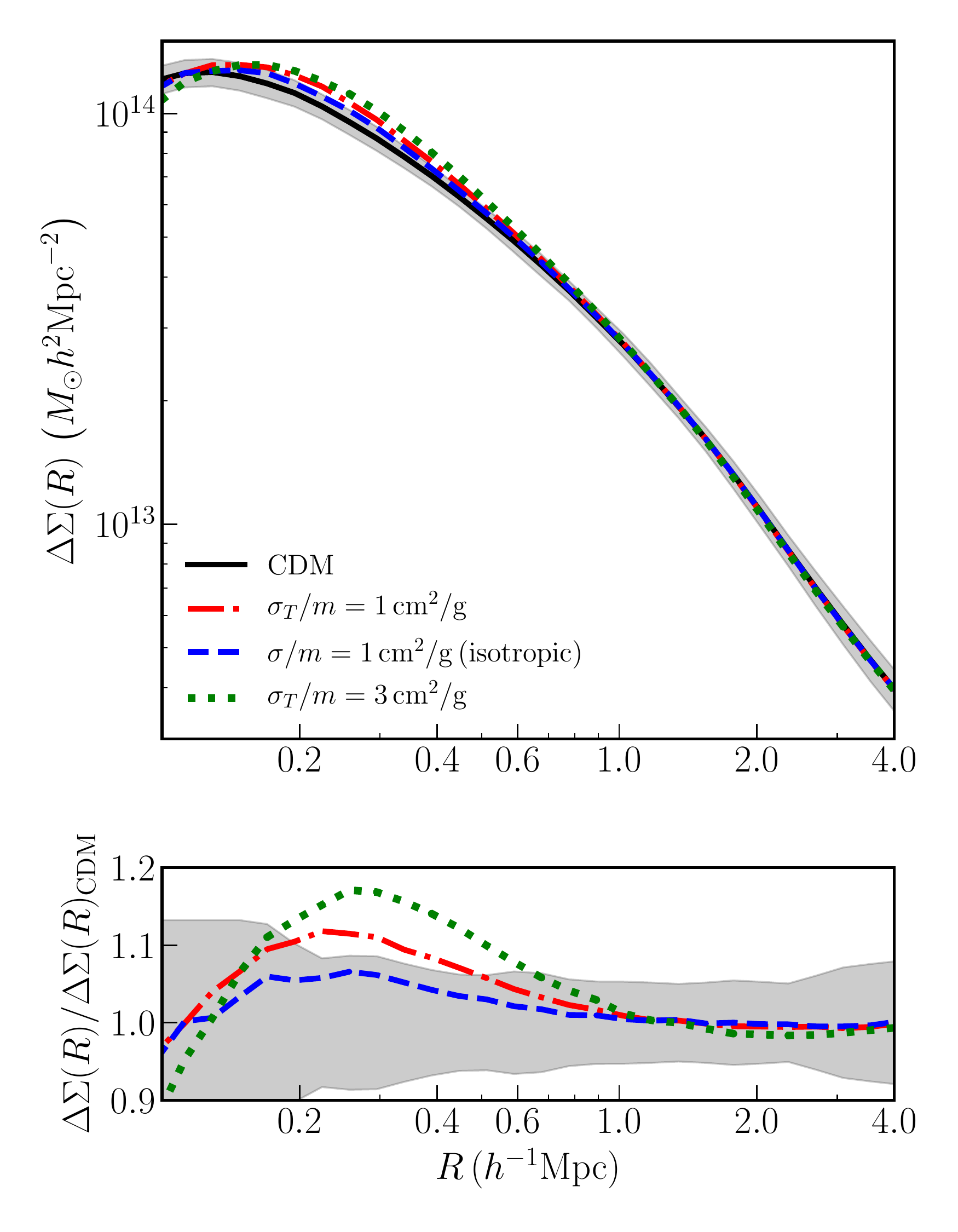}
\includegraphics[width=0.45\linewidth,]{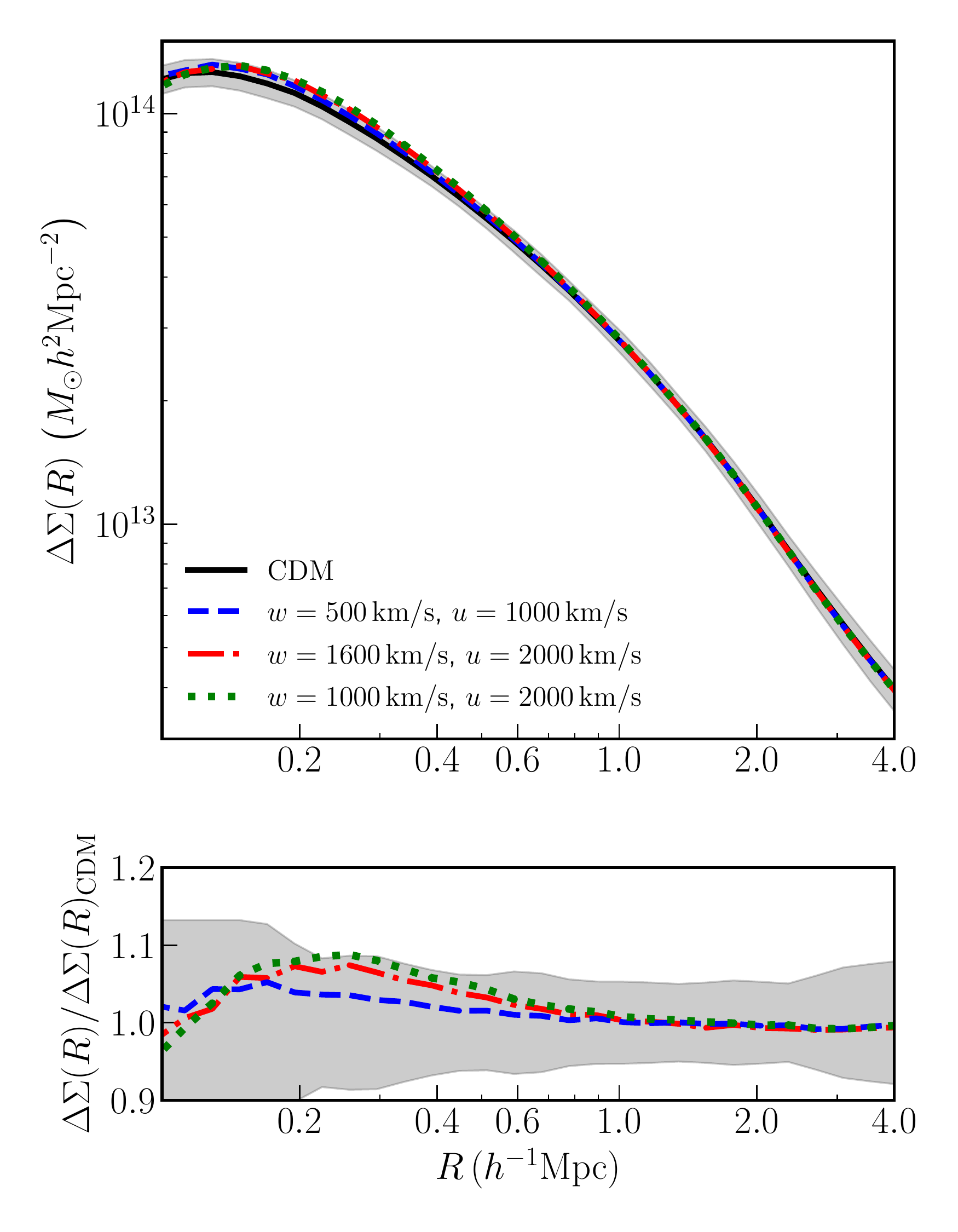}
\end{center}
\caption{Top: Stacked $\Delta \Sigma$ profile around halos in the mass range $1-2\times 10^{14}\hMsun$ in the velocity dependent (left panel) and velocity-independent (right panel) SIDM scenarios. Bottom: Ratio of the stacked $\Delta \Sigma$ profile in each self-interaction scenario to the profile in the CDM case. The shaded regions represent the approximate measurement error bars on $\Delta \Sigma$ from DES Y1 data. The left hand panels are for velocity-independent interactions, while the right hand panels are for velocity-dependent interactions.}
\label{fig:2d_profile_particles}
\end{figure}

%----------------------------------------------------------
\subsection{Projected distributions of matter and subhalos and comparison with data}
%----------------------------------------------------------

In the previous sections we concentrated on the stacked 3D density profiles around halos. Large-scale imaging surveys like the DES\footnote{\url{https://www.darkenergysurvey.org/}}, the HSC survey\footnote{\url{https://www.naoj.org/Projects/HSC/}}, and the LSST\footnote{\url{https://www.lsst.org}}  provide thousands of optically selected galaxy clusters that enable us to study the projected density profiles of matter and substructure in halos  using weak lensing measurements and cluster-galaxy cross-correlations respectively. The observable in the weak lensing measurements from these photometric surveys is the projected distribution of matter given by $\Delta \Sigma = \Sigma(< R)-\Sigma(R)$, where $\Sigma$ is the projected 2-d mass density profile, and $R$ is the projected distance from the cluster center. While the measurement of $\Delta \Sigma$ for a single cluster is noisy, stacking sufficient numbers of clusters beats down the noise, and we can obtain high signal to noise on the stacked measurement.

Fig. \ref{fig:2d_profile_particles} shows the stacked $\Delta \Sigma$ profiles for clusters in the mass range $1-2\times 10^{14}\hMsun$ from the CDM simulation along with the different models of self-interaction investigated in this paper. The shaded regions on each panel represent the measurement error on the projected matter distribution, $\Delta \Sigma$, from the first year data of the Dark Energy Survey (DES-Y1) adopted from \cite{Chang:2017hjt}. The errors from the weak lensing measurements is of the order of $\lesssim 10\%$ on the scales of interest. These measurements are already of the order of the signal generated by large cross sections (e.g., $\sigma_T/m > 3\,$cm$^2/$g). It is interesting to note that the current statistical constraints from weak lensing profiles are already quite close to the constraints on isotropic scattering cross-sections from the Bullet cluster which is $\sigma/m \lesssim 2\,$cm$^2/$g \cite{Robertson:2016xjh}. Year 3 data from DES (DES-Y3), has $3$ times as many clusters as the sample used in \cite{Chang:2017hjt}, and so the statistical error bars are expected to go down by a factor
of $\sqrt{3}$. For LSST, which will cover half the sky, and can go to higher redshifts, the statistical error bars on the stacked weak lensing profiles are expected to be as low as $2\%$ on these scales. This means that it may be possible to probe cross sections even smaller than $\sigma/m=1\,{\rm cm^2/g}$, denoted by the blue curve in Fig. \ref{fig:2d_profile_particles}. 
%We also point out that the distinctive behavior of the  density profiles in the case of self-interacting dark matter, where it is larger than the CDM profile for a range of radius but eventually turns over at smaller radii cannot be mimicked by a sample of halos which have a larger concentration. 
The profile of slopes  shown in Fig. \ref{fig:halo_profile_particles} is  significantly different from objects with higher concentrations (see \cite{Diemer:2014xya}), so if other systematics (like miscentering, for example) can be independently constrained, then upcoming weak lensing measurements may provide a competitive avenue for probing DM interactions.

It is also possible to place constraints on the strength of self-interaction by using the stacked projected satellite counts around the clusters, which has also been measured in \cite{Chang:2017hjt,Baxter:2017csy,Shin:2018pic, Zuercher:2018prq}. In fact this measurement has much higher statistical signal to noise compared to the lensing measurements referred to above. However, there are systematic uncertainties in interpreting these measurements with simulations, arising from the uncertainty in the recipe for matching subhalos from simulations to the observed satellite galaxies. It should be noted that, in fact, galaxy number density profiles from observations follow the particle distributions from simulations more closely than subhalo distributions (see Fig. 4 in \cite{Chang:2017hjt}). This can be attributed to tidal stripping (both real and artificial) of subhalos in the simulation causing subhalos to pass below the mass-resolution. Galaxies being more compact are not as affected by tidal disruptions. If we assume that the particle profile can be used directly to compare to the observed galaxy profiles, one could place stronger constraints on cross sections as low as $\sigma/m=2\,{\rm cm^2/g}$ or $\sigmaT/m=1\,{\rm cm^2/g}$.

%------------------------------------------------
\subsection{Shape of halo as a function of scale}
%------------------------------------------------

We investigate how the shapes of dark matter halos change as a function of radius from the halo center in the presence of self-interactions. To do this, we once again use all halos in the simulations in the mass range  $1-2\times 10^{14} \hMsun$. For each halo, we bin the particles associated with the halos in spherical shells, and then for each shell we compute the unweighted quadrupole tensor,
\begin{equation}
I^\alpha_{\mu\nu}= \sum_{i} x_{i,\mu} x_{i,\nu} \, ,
\label{eq:tensor}
\end{equation}
where the sum over $i$ represents the sum over all particles in a spherical shell labeled $\alpha$, while $\mu$, $\nu$ represent the axes. We then compute the square root of the eigenvalues of $I^\alpha_{\mu\nu}$, denoted by $\lambda_1$, $\lambda_2$, and $\lambda_3$, with $\lambda_1$ being the square root of the largest eigenvalue and $\lambda_3$ being similarly defined for the smallest eigenvalue. We then store the axis ratios $\lambda_2/\lambda_1$ and $\lambda_3/\lambda_1$ and find the average of these axis ratios over all halos in the mass bin. Note that for this measurement, the individual axis ratios of each halo has to be estimated before averaging, so as not to wash out the ellipticity signal due to the random orientation of the halos.

\begin{figure}
\begin{center}
\includegraphics[width=0.45\linewidth]{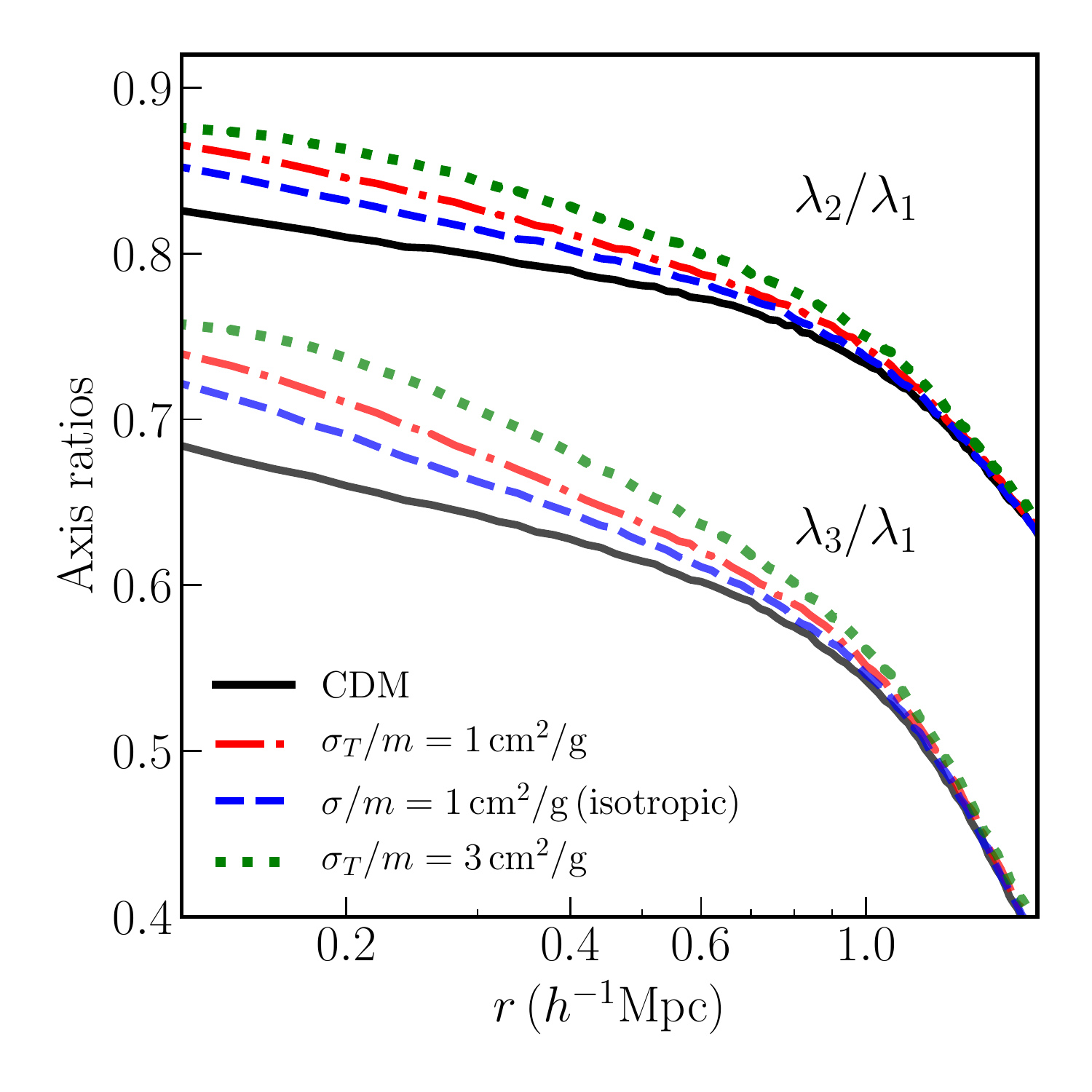}
\includegraphics[width=0.45\linewidth]{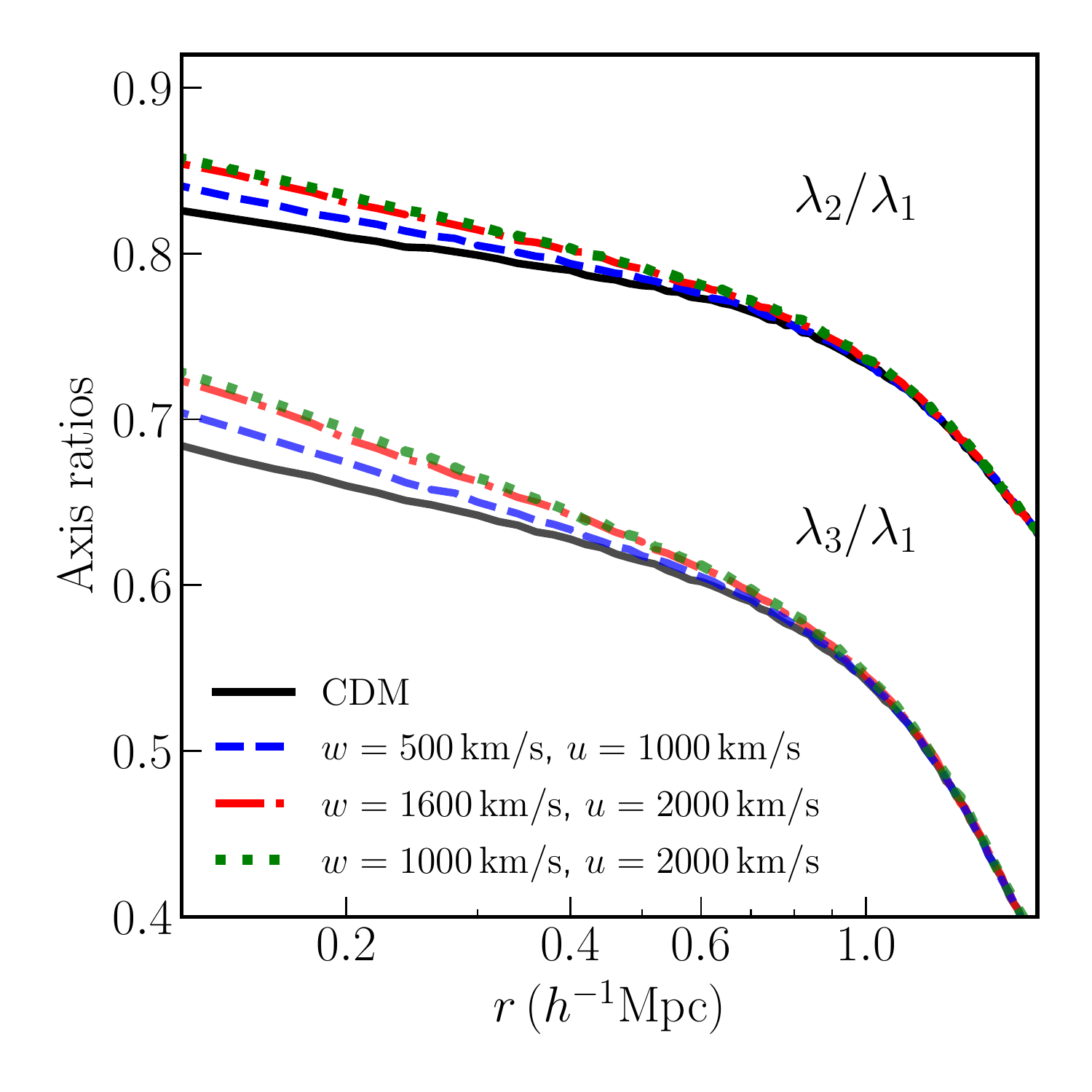}
\end{center}
\caption{Ratio of the square roots of the second largest to the largest eigenvalue ($\lambda_2/\lambda_1$) and smallest to largest eigenvalues ($\lambda_3/\lambda_1$) of the unweighted quadrupole tensor of halos (as defined in Equation \ref{eq:tensor}), averaged over all halos in the mass range $1-2\times 10^{14}\hMsun$. The left panel shows results for the velocity independent self-interactions, while the right panel shows the same profiles for the self-interaction models with velocity-dependent cross section. Halos become rounder in the presence of self interactions.}
\label{fig:ellip}
\end{figure}

\iffalse
\begin{figure}
\begin{center}
\includegraphics[width=0.45\linewidth]{radial_orbits_vel_ind}
\includegraphics[width=0.45\linewidth
]{radial_orbits_vel_dep}
\end{center}
\caption{Fraction of outgoing orbits that are radial (as defined in Equation \ref{eq:angle_of_orbit}) in the velocity-independent (left panel) and velocity-dependent (right panels) SIDM scenarios. In the presence of self-interactions, fewer particles move out from the center on radial orbits as compared to CDM, implying that self-interactions tend to scatter radial orbits into more circular orbits.
\neal{All this seems to show is that the velocity anisotropy gets smaller as the cross-section is increased.  To show that plunging orbits are depleted, I think you'd need to track actual particle orbits over time.}
}
\label{fig:radial_orbits}
\end{figure}
\fi

We plot the results of this measurement in Fig. \ref{fig:ellip}. Note that the axis ratios change rapidly at radii $\approx 1\hMpc$, just below the splashback radius, due to the transition from the relatively mixed inner regions and newly infalling matter \cite{Suto:2016,Snaith:2017}. The figure shows that self-interactions tend to increase the value of both $\lambda_2/\lambda_1$ and $\lambda_3/\lambda_1$ over the CDM case, meaning that the halos become rounder in the presence of the self-interactions. This effect has been seen previously for the isotropic interactions in \cite{Peter:2012jh}, but we find that this will be generally true irrespective of the details of the differential cross section. Note that the definition of the shape of halos used here is different from another halo shape definition used commonly in the literature, including \cite{Peter:2012jh} - the latter uses the ratios of the eigenvalues of the weighted moment of inertia tensor \cite{Allgood:2005eu} to define the shape. To confirm that our inference of halos in the chosen mass range getting rounder in the presence of self-interactions does not depend on the exact definition of shape, we compute the mean of the axis ratios of the chosen halos as taken from the \textsc{Rockstar} halo catalog. \textsc{Rockstar} computes these axis ratios using the eigenvalues of the weighted moment of inertia tensor, but only for the full halo, not as a function of radius. However, here too, we find that the axis ratios move closer to unity as the strength of the self-interactions is increased, supporting the conclusion that halo shapes become more spherical as particles experience more scattering.

\section{Summary and Discussion}
\label{sec:summary}
%===============================

In this paper we have investigated the signatures of self-interacting dark matter on mass profiles and radial subhalo distribution in cluster-sized halos of mass $1\times 10^{14}\hMsun$ to $2\times 10^{14}\hMsun$  for different self-interaction scenarios, including models with velocity-dependent and angle-dependent differential cross sections. We have implemented these models in the cosmological $N$-body code \textsc{Gadget-2} and used it to run large volume cosmological simulations of each of these models. 

With regard to the mass distribution within cluster halos, even though the interaction history of an individual particle is different for different self-interaction simulations, we find that the macroscopic signatures on the dark matter density profile are qualitatively similar on the scales we have considered, i.e. the primary effects on the matter distribution are a change of the density profile near the central regions and sphericalization of halo shapes at radii of $\lesssim 1\,\hMpc$. For velocity-independent interactions, these effects scale directly with the cross section, parameterized by $\sigma/m$ for isotropic interactions, and $\sigmaT/m$ for anisotropic interactions. We also find no qualitatively different signatures between the isotropic and anisotropic interactions we consider. For velocity-dependent interactions, we find that the overall effects are smaller than the velocity-independent ones considered in this paper. 

One of the most important results of this investigation is that the halo profile is affected by self-interactions at an appreciable level out to projected distance $\sim 0.7\, \hMpc$, even for interaction strengths that are not completely ruled out by e.g. the Bullet Cluster measurements \cite{Markevitch:2003at,Robertson:2016xjh}. This effect can be probed using the stacked $\Delta \Sigma$ profiles of clusters, measured via weak lensing. In fact, such measurements around similar mass objects have been performed already \cite{Chang:2017hjt}.  Current uncertainties are  already at a level sufficient to probe interactions with $\sigmaT\geq 3\,{\rm cm^2/g}$. As ongoing and future surveys find larger numbers of clusters, either using optical selections or the Sunyaev-Zeldovich effect \cite{Carlstrom:2002na}, the error bars on the lensing profiles are expected to shrink rapidly. Just going from DES Y1 data to Y3 (Year 3) data, the number of clusters is expected to increase by a factor of 3, providing significantly tighter constraints on the self-interaction cross sections. In the future, LSST, with larger sky coverage ($f_{\rm sky} \sim 0.5$) and greater depth is expected to measure the lensing profile on relevant scales at $\sim 2\%$ level. If this target is met, and potenital systematics like miscentering can be mitigated, it is likely that we achieve sufficient statistical power to rule out self-interaction cross sections even below $\sigma/m <1\,{\rm cm^2/g}$, or equivalently $\sigmaT/m <0.5\,{\rm cm^2/g}$. The fact that measurements in the outskirts of clusters can be used in studies of dark matter self-interactions is important for the following reason: even though the strongest signatures from self-interactions are expected within the scale radius of these clusters ($\lesssim 0.15\, \hMpc$), the baryons contribute significantly to the total mass within this region and effects of feedback can substantially modify the mass distribution within it, increasing systematic uncertainty of any measurement.  Furthermore, the light of the brightest cluster galaxy (BCG) obscures background galaxies and member galaxies in the innermost regions making it difficult to distortions of source very close to the cluster center along the line of sight. 

We have also focused on the stacked distribution of subhalos within galaxy clusters. The subhalos are tracers of the dark matter potential and are the visible component of clusters, lit up by the galaxies within them. The imaging surveys mentioned above also observe the number density profiles of substructure around the clusters, and such profiles have already been measured with high accuracy. We find that the number density of subhalos is suppressed in most of the models of self-interactions that we have investigated in this paper. The difference can be as large as $10\%$ out to the halo boundary of these clusters for anisotropic interactions with $\sigmaT/m=1\,{\rm cm^2/g}$. Again, this suggests that SIDM cross sections can potentially be constrained using satellite counts in the outskirts of massive clusters. Even though the statistical signal to noise of the satellite counts is much higher than weak lensing measurements, the biggest systematic to using satellite counts to constrain self-interaction cross sections is a proper understanding of the mapping between subhalos in simulations and the observed satellite galaxies. If this systematic can be dealt with, one can expect even stronger constraints on the self-interactions than that coming from weak lensing in a survey like LSST.

We find that the subhalo distribution can encode signatures about the velocity dependent nature of the interaction cross-section. In hierarchical structure formation, the halos that will eventually merge to form subhalos within cluster mass halos have internal velocity dispersions that are smaller than the host, and can have significantly different interaction cross-sections within them. This alters the structure of subhalos making them more susceptible to being disrupted by both tidal forces and SIDM interactions, leading to a suppression of the subhalo density profile, even when the cross-section at the relative velocity scale between the host and the subhalo is of the order of $1 \mathrm{cm^2/g}$. While this feature potentially leads to a degeneracy in model predictions when the subhalo profile is the only observable, the degeneracy can be broken by the measurement of the lensing profile, which is sensitive to the relative velocity scale only.

Apart from the density profiles of particles and subhalos, we have also explored signatures of SIDM on the derivative of the density profiles. In particular we focused on the splashback radius which forms the phase space boundary of the halo, and corresponds to the minimum of the logarithmic slope of the profile. The splashback radius for dark matter particles can be probed by stacked lensing profiles around galaxy clusters. We find that when clusters that form early (or have high concentration for the CDM counterpart) are stacked together, the location of the particle splashback is shifted to a smaller radius in the presence of self-interactions, while late forming halos do not show an analogous shift. Stacking halos in observations based on properties that correlate with age, like the magnitude-gap for example could, in principle, be used to isolate this effect, where the signal would manifest as a larger difference between the  splashback of early and late forming halos in a self-interacting scenario, when compared to the standard CDM expectation. 

 For subhalos we find that splashback radius moves to a smaller value by about $20\%$ when the cross-section is as high as $\sigmaT/m=3\,{\rm cm^2/g}$. This happens because at such high cross sections, subhalos on radial orbits, which would otherwise have had the largest apocenters, are preferentially destroyed on their very first passage close to the host center. However, these cross-sections correspond to an isotropic $\sigma/m\sim6\,{\rm cm^2/g}$ which are ruled out by Bullet cluster constraints. We note that the depth of the minimum of the slope profile in models with self-interacting dark matter begins to deviate significantly from the CDM model, especially for the subhalo number density profile -- the slope at the splashback radius becomes shallower for the strongest anisotropic interaction cross-section. While most current observations focus only on determining the splashback radius, it is also, in principle, possible to measure the depth of the dip of the slope profile around the splashback radius, as accuracy of such measurements improves.

Wide and deep optical and CMB surveys give us an unparalleled statistical sample of galaxy clusters. The number of clusters in the DES alone will increase by a factor of 3, while the HSC survey will provide deeper data. At the same time, the number of clusters in SZ and X-Ray selected samples is expected to rise to $\sim10^5$. Precision measurements of the radial distribution of mass and satellite galaxies around galaxy clusters will, therefore, help to place strong constraints on self-interaction cross-sections that will be complementary to other probes of such interactions, such as cluster mergers \cite{Markevitch:2003at,Robertson:2016xjh} or BCG displacements \cite{Kim:2016ujt,Harvey:2018uwf}.  

%=========================
\acknowledgments

We thank Annika Peter and Stacy Kim for help with code comparison, and Chihway Chang and Eric Baxter for providing the measurement errors from data. We thank Annika Peter, Felix Kahlhoefer, Tom Abel, Ethan Nadler and Bhuvnesh Jain for helpful discussions. This work used the Sherlock cluster at the Stanford Research Computing Center, as well as the Midway cluster at the University of Chicago Research Computing Center.  Research at Perimeter Institute is supported in part by the Government of Canada through the Department of Innovation, Science and Economic Development Canada and by the Province of Ontario through the Ministry of Economic Development, Job Creation and Trade. AK was supported by the NSF grant AST-1714658 and by the Kavli Institute for Cosmological Physics at the University of Chicago through grant PHY-1125897 and an endowment from the Kavli Foundation and its founder, Fred Kavli.

\bibliography{sidm}
\bibliographystyle{JHEP}

%\appendix

\end{document}